\documentclass[a4paper,fleqn,usenatbib]{mnras}

\usepackage{newtxtext,newtxmath}

\usepackage{amsmath}
\usepackage{physics}
\usepackage[T1]{fontenc}
\DeclareRobustCommand{\VAN}[3]{#2}
\let\VANthebibliography\thebibliography
\def\thebibliography{\DeclareRobustCommand{\VAN}[3]{##3}\VANthebibliography}

\usepackage{graphicx}	
\usepackage{amsmath}	
\usepackage{amssymb}	
\usepackage{multicol}

\title[Transport and stability in accretion discs]{Angular momentum transport and thermal stabilization of optically thin, advective accretion flows through large-scale magnetic fields}

\author[Datta, Mondal \& Mukhopadhyay]{
Sudeb Ranjan Datta,$^{1}$\thanks{E-mail: sudebd@iisc.ac.in}, Tushar Mondal$^{2}$, Banibrata Mukhopadhyay$^{1}$
\\
$^{1}$Department of Physics, Indian Institute of Science, Bangalore-560012, India\\
$^{2}$International Centre for Theoretical Sciences, Tata Institute of Fundamental Research, Bangalore
560089, India
}

\date{Accepted XXX. Received YYY; in original form ZZZ}

\pubyear{2021}

\begin{document}
\label{firstpage}
\pagerange{\pageref{firstpage}--\pageref{lastpage}}
\maketitle

\begin{abstract}
Outward transport of angular momentum, as well as viscous and thermal stability, are the necessary criteria for the formation of accretion disc and to radiate steadily. Turbulent motions originating from magneto-rotational instability or hydrodynamic instability can do the required transport. We explore the effect of a large-scale magnetic field (LSMF) over the turbulent transport in an optically thin advective disc. In this work, turbulent transport is represented through the usual Shakura-Sunyaev $\alpha$-viscosity. The evolution of the magnetic field and other variables is found by solving vertically integrated height averaged magnetohydrodynamic equations. Depending on its configuration, the LSMF can support or oppose $\alpha$ in outward transport of angular momentum. Once outward transport of angular momentum is assured, i.e., formation of the disc is confirmed through the combined effect of $\alpha$-viscosity and the LSMF, we explore the impact of the LSMF in thermally stabilizing the disc. As found earlier, we also find that the advection of heat energy becomes zero or negative with increasing accretion rate. That is why at or above a critical accretion rate, the optically thin advective disc becomes thermally unstable. We show however that with the addition of a strong enough magnetic field, the disc regains its thermal stability and Joule heating turns out to play the key role in that. Throughout our analysis the plasma-$\beta$ ($\beta_\mathrm{m}$) remains within the range of 5-$10^3$, which does not impose any restriction in the simultaneous operation of the LSMF and the turbulent transport.

\end{abstract}




\section{Introduction}

Outward transportation of angular momentum is the necessary requirement for the formation of accretion disc in a low mass X-ray binary where matter flows from the binary companion to the compact object through Roche lobe overflow. Molecular viscosity is too inefficient to meet the requirement of necessary transport (section 4.7 of \citealt{Frank2002}). \cite{Shakura1973} introduced the idea of $\alpha$-viscosity: viscous stress tensor $\boldsymbol{\sigma_\mathrm{\bf{ik}}^{'}}$ is assumed to scale with local gas pressure and $\alpha$ is the constant of proportionality. In that work itself, they invoked the turbulent, chaotic motion of matter and/or magnetic field as the possible reason for the origin of $\alpha$. Till now, many works have focused on understanding the origin of turbulent transport, which possibly gives rise to $\alpha$. Most promising is the magneto-rotational instability (MRI: \citealt{Velikhov1959, Chandrasekhar1960, Balbus1998}). A few works (\citealt{Ghosh2021} and references therein) showed that hydrodynamic instability could also be a plausible reason if some extra force is applied to the system.

However, some other works explored the role of the large-scale magnetic field (LSMF) in the transportation of angular momentum. Mostly they focused on the launching of outflows and jets from the disc with the help of the LSMF, and the outflows or winds (least collimated outflow) help in transporting angular momentum outward as they leave the disc. \cite{Zhu2018}, \cite{Mishra2020} and \cite{Jacquemin-Ide2021} did the global simulation of the geometrically thin turbulent accretion disc in the presence of a net vertical magnetic field. The whole region of the flow is being divided vertically into regions with separated dominance of turbulent and laminar (LSMF) transports. However, these works assumed the presence of a vertical magnetic field to focus on the launching of winds and showed that wind does not help significantly in transporting angular momentum outward and in the accretion of the matter. Most of the accretion occurs through the coronal atmosphere where $\beta_\mathrm{m} < 1$, i.e., the magnetically dominated region which lies above the turbulent disc. It is found that radial laminar torque associated with the LSMF can play a crucial role in angular momentum budget. \cite{Mukhopadhyay2015} explored the role of the LSMF in transporting angular momentum in an optically thin advective accretion disc, and without even the presence of $\alpha$-viscosity, the disc was shown to be formed. This result and the simulations are quite encouraging in establishing a key role of the LSMF in transport.

We explore the contribution of the LSMF in transporting angular momentum in optically thin advective accretion disc along with turbulent transport (approximated through $\alpha$-prescription) in this work. \cite{Jacquemin-Ide2021} nicely separated out the contribution of the laminar term (or LSMF or mean-field) and turbulent part (fluctuations of velocity and magnetic field) in angular momentum transport. To simplify the situation and focus on the LSMF, we approximate the turbulent transport through $\alpha$-prescription and solve for the evolution of the magnetic field and its effect strictly by solving magnetohydrodynamic equations in pseudo-Newtonian potential.

Besides the transportation of angular momentum, the disc has to be viscously and thermally stable to exist in nature and give radiation steadily. We know that without advection, optically thin flows are, though viscously stable, thermally unstable (\citealt{Shapiro1976, Pringle1976, Piran1978, Abramowicz1995}). Optically thin flows lead to incomplete thermalization of ions and electrons, resulting in two temperature flows (\citealt{Narayan1995, Chakrabarti1995, Nakamura1997, Manmoto1997, Narayan1997, Mahadevan1997, Mandal2005, Rajesh2010}). In the absence of a magnetic field, viscous dissipation is the only source of heating. Ions gain most of the energy from gravitational potential through viscous dissipation and transfer it to electrons through Coulomb coupling ($Q^\mathrm{ie}$). Electrons are efficient radiators. They cool the system through radiation. Two temperature flow exists with ions and electrons at higher and lower temperatures respectively. A little increase in ion temperature leads to higher energy transfer to the electron. As electron cooling, which is through bremsstrahlung, synchrotron, and their comptonization, remains unaffected, electrons are heated up, and their temperature rises. Higher electron temperature decreases the Coulomb coupling $Q^\mathrm{ie}$, i.e., the energy transfer from the ion. This decrement in $Q^\mathrm{ie}$ is more than its initial increment. Then as a consequence, the ions transfer less energy; they are heated up more and increase its temperature further. In this way, without advection, the increment in ion temperature leads to their further increment in temperature, and that is why optically thin accretion flows are thermally unstable without advection (\citealt{Pringle1976}, section 3.3 of \citealt{DGK2007}). If the advection of heat is taken into account, it acts as a cooling mechanism and provides thermal stability to the flow. A few works had been done and, keeping in mind the stability, the accretion flows are categorized into different classes depending on mass accretion rate and optical thickness (\citealt{Chen1995}). The Coulomb coupling, which cools the ion, depends on the accretion rate steeply in comparison with viscous dissipation, which heats the ion. With increasing accretion rate, Coulomb coupling increases more than the viscous dissipation, leading to the decrement of advection. A critical value of accretion rate exists for which dissipated heat balances Coulomb coupling. At or above that critical accretion rate, advected energy through ions becomes zero or negative, leading to thermally unstable optically thin flow (\citealt{Narayan1996, Esin1997}). As the viscous dissipation increases proportionally with $\alpha$-viscosity, the critical accretion rate also depends on the value of $\alpha$. \cite{Narayan1996}, \cite{Esin1997} found this critical accretion rate to be $\sim 0.4\alpha^2\dot{M}_\mathrm{Edd}$. Remember that in general this $\alpha$-viscosity includes turbulent and the laminar contributions. In this work, we denote $\alpha$-viscosity as a presentation of turbulent contribution only. On a note of caution, this critical accretion rate depends on the $\delta$ parameter, the fraction of viscously dissipated heat that goes directly to the electron, as well as on the chosen radius value.

Above that critical accretion rate, it is believed that no optically thin stable accretion flow can exist. As cooling of ions is more significant than heating at a higher accretion rate, cooling instability kicks in, and it is believed that eventually, it will converge to an optically thick disc (\citealt{Mineshige1996, Machida2006}). \cite{Yuan2001} introduced the idea of luminous hot accretion flow (LHAF). It focuses on the increment of internal energy as the matter is accreted inward instead of the advection of heat energy. This relaxation gives the freedom to add heating due to compression with the viscous dissipation to balance the Coulomb coupling and keep the flow hot. Although this pushes the limit of accretion rate to a higher value, these flows become thermally unstable against local perturbations for higher accretion rates (\citealt{Yuan2003}). However, here we keep the positive value of advection of energy as the necessary criterion for the thermal stability of the optically thin accretion flow. We find that strong LSMF (weak enough to keep $\beta_\mathrm{m}$ in the range $\sim$ 5-$10^3$) also can push the accretion rate limit to a higher value. A few works (\citealt{Oda2007, Oda2009, Oda2010, Oda2012, Sadowski2016}) already showed that the LSMF could stabilize the disc. Most of the earlier analyses were local ones. Although \cite{Oda2012} did the global analysis for the optically thin disc as analyzed in this work, the Maxwell stress is assumed to be proportional to total pressure, and through this assumption, magnetic heating is also taken into account. Joule heating is not computed self-consistently in that work. \cite{Sadowski2016} showed through general relativistic magnetohydrodynamic simulation that a magnetic field with certain initial geometry can stabilize geometrically thin disc in which instability due to radiation pressure kicks in at a higher accretion rate. 

We find that depending on the configuration of the LSMF (relative strength and orientation of different magnetic field components), it can oppose or help $\alpha$-viscosity (turbulent contribution) in the outward transportation of angular momentum. Radial evolution of all the components of magnetic fields is taken care of self-consistently. We fix the configuration of the LSMF so that it helps $\alpha$-viscosity parameter the most in transporting angular momentum outward. With a weak vertical field, the suitable configuration turns out to be the toroidally dominated field which can efficiently transport angular momentum outward. As indicated above, a suitable configuration of the LSMF can oppose $\alpha$-viscosity too and prohibit the accretion flow from happening. In addition to transportation, the LSMF also makes the disc cooler, reduces the outward transport of angular momentum through turbulent $\alpha$-viscosity, even though the value of $\alpha$ remains the same. We find that the work done by the gas due to compression or expansion increases with the increase of the strength of the magnetic field, leading to the decrease in ion temperature. Moreover, we find that the LSMF helps in stabilizing the disc by increasing the total heating in the system through Joule heating. In this way, strong LSMF can transport angular momentum outward along with $\alpha$-viscosity and prohibit advective accretion disc with a larger accretion rate to become thermally unstable. Subsequently we also find that a strong vertical field also can transport angular momentum efficiently, and correspondingly it can stabilize the disc thermally like the toroidally dominated field. For all our investigated cases, $\beta_\mathrm{m}$ remains greater than unity always, which is quite encouraging as it does not restrict the simultaneous operation of the LSMF and the $\alpha$-viscosity even if we assume the source of $\alpha$ solely to be MRI (as MRI ceases for $\beta_\mathrm{m}<5$, \citealt{Balbus1991}).

We discuss the assumptions and magnetohydrodynamic equations, which we solve, in the next section. In section \ref{section Results}, we present our results: when the vertical field is weak, how different configurations of the LSMF oppose and help $\alpha$-viscosity, how the LSMF affects dynamics as well as different quantities related to the thermal balance of the disc. The thermal instability kicks in above a critical accretion rate for the non-magnetic or weakly magnetic case and the flow regains stability with strong LSMF, which are shown in sections \ref{section instability due to mdot} and \ref{section LSMF stabilizing disc} respectively. Outward transport of angular momentum through the strong vertical field is discussed in section \ref{section_vertical_field_transport}. After discussing a few crucial points in section \ref{section Discussion} and mentioning the caveats of our work in section \ref{section caveats}, we summarize at last in section \ref{section summary}. 

\section{Formalism}
\label{section Formalism}
\subsection{Fundamental equations}

We follow the standard method to find the magnetohydrodynamics of optically thin, two-temperature, advective accretion flows around black holes. The accreting gas under consideration consists of electrons and ions, and thus behaves as a two-temperature system, apart from radiation. The number densities of ions and electrons are equal by charge neutrality, i.e. $n_\mathrm{i} = n_\mathrm{e} =n$. Here, we adopt cylindrical co-ordinates $(r, \phi, z)$ and solve the following equations to find the magnetohydrodynamic solutions. The mass continuity equation and the momentum balance equation in the presence of magnetic fields are, respectively,
\begin{equation}\label{mass continuity}
\frac{\partial \rho}{\partial t}+\nabla.(\rho \boldsymbol{v})=0,
\end{equation}
\begin{equation}\label{MHD equation}
\frac{\partial v_\mathrm{i}}{\partial t}+(v_\mathrm{j}\partial_\mathrm{j})v_\mathrm{i}=F_\mathrm{i}-\frac{1}{\rho}\partial_\mathrm{i} p+\frac{1}{\rho}\partial_\mathrm{k}\boldsymbol{\sigma_\mathrm{\bf{ik}}^{'}}+\left[\frac{1}{4\pi\rho} (\nabla\times\boldsymbol{B})\times\boldsymbol{B}\right]_\mathrm{i},
\end{equation}
where $\rho$ is the density of the flow, $\boldsymbol{v}$ is the velocity vector, $\vb*{B}$ is the magnetic field vector, $\boldsymbol{\sigma}_\mathrm{ik}^{'}$ is the viscous stress tensor which appears due to turbulent viscosity, $p$ is flow pressure which includes gas (of ion and electron) and radiation. Note that $p$ does not include the magnetic pressure and takes the form
\begin{equation}\label{eq:pressure}
p=p_\mathrm{i}+p_\mathrm{e}+p_\mathrm{rad}=\frac{\rho kT_\mathrm{i}}{\mu_\mathrm{i}m_\mathrm{p}}+\frac{\rho kT_\mathrm{e}}{\mu_\mathrm{e}m_\mathrm{p}}+\frac{1}{3}aT_\mathrm{eff}^4,
\end{equation}
where $k$ is the Boltzmann constant, $a$ is the radiation constant which is related to the Stefan's constant $\sigma_\mathrm{s}$ through $a=4\sigma_\mathrm{s}/c$, $m_\mathrm{p}$ is the proton mass, $c$ is the speed of light. $\mu_\mathrm{i}$ and $\mu_\mathrm{e}$, respectively, are the effective molecular weights for ions and electrons, $T_\mathrm{i}$ and $T_\mathrm{e}$, respectively, are ion and electron temperatures, and $T_\mathrm{eff}$ is the effective surface temperature. Following \cite{Narayan1995}, $T_\mathrm{eff}$ is calculated such that $\sigma_\mathrm{s}T_\mathrm{eff}^4$ gives the flux emitted from the disc. $F_\mathrm{i}$'s are the different components of gravitational pseudo-Newtonian force at the equatorial plane of the disc. The radial component of the force is given by \cite{Mukhopadhyay2002} as
\begin{equation}
\label{equation F(r)}
F_\mathrm{r}(r)=-\frac{(r^2-2\mathrm{a}\sqrt{r}+\mathrm{a}^2)^2}{r^3(\sqrt{r}(r-2)+\mathrm{a})^2}, 
\end{equation}
where throughout in our calculation, the Kerr parameter $\mathrm{a}=0$ as for the non-rotating black hole (same as \citealt{Paczynski1980}). 

The energy balance equations for ions and electrons are, respectively,
\begin{equation}\label{ion TdS}
Q_\mathrm{adv, i}=(1-\delta)Q^\mathrm{+}-Q^\mathrm{ie},
\end{equation}
\begin{equation}\label{electron TdS}
Q_\mathrm{adv, e}=\delta Q^\mathrm{+}+Q^\mathrm{ie}-Q^\mathrm{rad},
\end{equation}
where $Q^\mathrm{+}$ is the heating rate per unit volume, $Q^\mathrm{ie}$ is the Coulomb coupling term through which ions transfer energy to electrons per unit volume per unit time, $Q^\mathrm{rad}$ is the radiative cooling rate per unit volume, and $\delta$ is the fraction of heating which directly goes to electrons. $\delta$ and $T_\mathrm{e}$ play a degenerate role to compute the disc spectra. As here we are not focusing on the spectra emitted from the disc, we keep $\delta=0$ throughout this work. $Q_\mathrm{adv, i}$ and $Q_\mathrm{adv, e}$ denote the advected heat per unit volume per unit time by ions and electrons, respectively. As the heating and cooling of both ions and electrons differ, a substantial fraction of the dissipated heat is stored as entropy differently in ions and electrons. Due to significant radial inward velocity, the stored heat is advected inward. These can be expressed as
\begin{equation*}
Q_\mathrm{adv, i}=-v_\mathrm{r}T_\mathrm{i} \frac{dS}{dr}\bigg|_\mathrm{ion},\ \text{and}\  Q_\mathrm{adv, e}=-v_\mathrm{r}T_\mathrm{e} \frac{dS}{dr}\bigg|_\mathrm{electron},
\end{equation*}
where $S$ is the specific entropy, and $v_\mathrm{r}$ is the radial velocity. The negative sign appears naturally because the value of $v_\mathrm{r}$ is negative in accretion flows. $Q^\mathrm{+}$ consists of both the viscous and magnetic dissipation parts, as given by 
\begin{equation*}
	Q^\mathrm{+}=Q^\mathrm{vis}+Q^\mathrm{mag}=\frac{\boldsymbol{\sigma_\mathrm{\bf{ik}}^{'}}^2}{\eta_\mathrm{V}}+\frac{j^2}{\sigma},
\end{equation*}\label{heating}
where the first term in the right-hand side is due to viscous dissipation (turbulent contribution in heating), and the second term is due to Joule heating (magnetic contribution). Following the standard notations, here, $\eta_\mathrm{V}$ is dynamic viscosity, $\boldsymbol{j}=$ (c/4$\pi$) $\nabla\times\boldsymbol{B}$ is the current density, and $\sigma$ is the conductivity. The Coulomb coupling term behaves as the heating term for electrons. It is given by \cite{Stepney1983} as
\begin{multline*}
	Q^\mathrm{ie}=\frac{3}{2}\frac{m_\mathrm{e}}{m_\mathrm{p}}n_\mathrm{e}n_\mathrm{i}\sigma_\mathrm{T}c\frac{kT_\mathrm{i}-kT_\mathrm{e}}{K_2(1/\theta_\mathrm{e})K_2(1/\theta_\mathrm{i})}ln\Lambda\\
	\times\Big[\frac{2(\theta_\mathrm{e}+\theta_\mathrm{i})^2+1}{(\theta_\mathrm{e}+\theta_\mathrm{i})}K_1\Big(\frac{\theta_\mathrm{e}+\theta_\mathrm{i}}{\theta_\mathrm{e}\theta_\mathrm{i}}\Big)+2K_0\Big(\frac{\theta_\mathrm{e}+\theta_\mathrm{i}}{\theta_\mathrm{e}\theta_\mathrm{i}}\Big)\Big]\ \text{ergs  cm}^{-3} \text{s}^{-1},
\end{multline*}
where $m_\mathrm{e}$ is the electron mass, $\sigma_\mathrm{T}$ is the Thomson scattering cross-section, $K$'s are modified Bessel functions, $ln\Lambda$ is the Coulomb logarithm (roughly $ln\Lambda \sim 20$), and the dimensionless electron and ion temperatures are defined by, respectively, 
\begin{equation*}
	\theta_\mathrm{e}=\frac{kT_\mathrm{e}}{m_\mathrm{e}c^2},\ \text{and}\ \theta_\mathrm{i}=\frac{kT_\mathrm{i}}{m_\mathrm{p}c^2}.
\end{equation*} 
In the calculation of $Q^\mathrm{ie}$, \cite{Stepney1983} assumed all the ions are protons. For a more general case, \cite{Narayan1995} introduced one numerical correction factor of 1.25 if we assume 75\% H and 25\% He instead of assuming all ions are protons. Also, for technical reasons, as \cite{Oda2010} approximated, we use the following formula of Coulomb coupling for our calculation,
\begin{equation}
Q^\mathrm{ie}=5.61\times10^{-32}n_\mathrm{e}n_\mathrm{i}(T_\mathrm{i}-T_\mathrm{e})\frac{\sqrt{2\pi}+\sqrt{\theta_\mathrm{e}+\theta_\mathrm{i}}}{\theta_\mathrm{e}+\theta_\mathrm{i}}\ \text{ergs  cm}^{-3} \text{s}^{-1}.
\end{equation}
Interestingly, it uses no special functions and is accurate to within a factor of $2$ when $\theta_\mathrm{i}<$ 0.2 (\citealt{Dermer1991}). For all our investigated cases $\theta_\mathrm{i}$ remains $<0.2$.

Regarding the radiative cooling, we consider that the electrons cool via different cooling processes: bremsstrahlung ($Q^\mathrm{Br}$), synchrotron ($Q^\mathrm{Sy}$), and the inverse comptonization processes of bremsstrahlung radiation ($Q^\mathrm{BrC}$), as well as synchrotron soft photons ($Q^\mathrm{SyC}$). Hence, the radiative cooling rate per unit volume can be expressed as
\begin{equation*}\label{cooling}
	Q^\mathrm{rad}=Q^\mathrm{Br}+Q^\mathrm{Sy}+Q^\mathrm{BrC}+Q^\mathrm{SyC}.
\end{equation*}
We follow \cite{Narayan1995} to formulate all the cooling processes. Here we do not rewrite the expression for each cooling as there is no explicit requirement. However, it is worth mentioning the exact expression for synchrotron cooling because there is a cutoff frequency ($\nu_\mathrm{c}$) below which it is self-absorbed, and we need to find $\nu_\mathrm{c}$ at each radius self-consistently to estimate synchrotron as well as total cooling from the disc. The expression for $\nu_\mathrm{c}$ is given by
\begin{equation*}
	\nu_\mathrm{c}=\frac{3}{2}\nu_\mathrm{0}\theta_\mathrm{e}^2x_\mathrm{M},\ \text{with}\  \nu_\mathrm{0}=2.8\times10^6 |\boldsymbol{B}| \ \text{ Hz},
\end{equation*}
where |$\boldsymbol{B}$| is expressed in Gauss. To find the parameter $x_\mathrm{M}$, we numerically solve the following transcendental equation for $x_\mathrm{M}$ at each radius $R$,
\begin{multline}\label{transcendental equation xM}
exp(1.8899x_\mathrm{M}^{1/3})=2.49\times10^{-10}\frac{4\pi n_\mathrm{e}R}{|\boldsymbol{B}|}\frac{1}{\theta_\mathrm{e}^3K_2(1/\theta_\mathrm{e})}\\\times\Big(\frac{1}{x_\mathrm{M}^{7/6}}+\frac{0.40}{x_\mathrm{M}^{17/12}}+\frac{0.5316}{x_\mathrm{M}^{5/3}}\Big),
\end{multline}
which is actually the equation (3.14) of \cite{Narayan1995}. Solving this transcendental equation at each radius gives $x_\mathrm{M}$ and consequently $\nu_\mathrm{c}$. Finally to estimate synchrotron cooling we adopt local approximation following \cite{Narayan1995}, as given by
\begin{equation*}
	Q^\mathrm{Sy}\approx\frac{2\pi}{3c^2}kT_\mathrm{e}(R)\frac{\nu_\mathrm{c}^3(R)}{R}\ \text{ergs  cm}^{-3} \text{s}^{-1}.
\end{equation*}
This completes the formalism part corresponding to the energy balance. 

The other two fundamental equations for magnetohydrodynamics are the induction equation and the equation for no magnetic monopole. These are, respectively,
\begin{equation}\label{induction equation}
\frac{\partial \boldsymbol{B}}{\partial t}=\nabla\times(\boldsymbol{v}\times\boldsymbol{B})+\eta_\mathrm{B}\nabla^2\boldsymbol{B},
\end{equation}
\begin{equation}\label{no monopole}
\nabla.\boldsymbol{B}=0,
\end{equation}
where $\eta_\mathrm{B}=c^2/(4\pi\sigma)$ is the magnetic diffusivity. We relate $\eta_\mathrm{B}$ to kinematic viscosity ($\nu_\mathrm{V}=\eta_\mathrm{V}/\rho$) by using the magnetic Prandtl number, $Pr=\nu_\mathrm{V}$/$\eta_\mathrm{B}$. As the turbulence can not propagate supersonically and the eddy size can not be larger than the scale-height of the disc, the turbulent kinematic viscosity can be expressed in terms of $\alpha$-viscosity as given by 
\begin{equation*}
\nu_\mathrm{V}=\alpha c_\mathrm{s} H\approx\alpha\sqrt{\frac{p}{\rho}}H,
\end{equation*}
which finally leads to 
\begin{equation*}
	\eta_\mathrm{B}=\frac{1}{Pr}\Big(\alpha\sqrt{\frac{p}{\rho}}H\Big),
\end{equation*}
where $c_\mathrm{s}$ represents the sound speed. Recent global simulation (\citealt{Zhu2018}) has reported the value of $Pr$ to be of the order of unity. We fix $Pr=1$ for our entire calculation.

\subsection{Assumptions over fundamental equations}
\label{section assumptions over fundamental equations}
We assume the vertical velocity of the flow to be zero ($v_\mathrm{z}=0$) to focus only on the disc dynamics, not on the wind or outflow. As only 5\% of the angular momentum is transported by wind (\citealt{Zhu2018}), ignoring $v_\mathrm{z}$ will not affect the angular momentum budget significantly. We also assume steady-state condition ($\partial/\partial t=0$), azimuthally symmetric ($\partial/\partial \phi=0$) flow to make the problem simpler and doable. As usual, we take into account only $\sigma_\mathrm{r\phi}^{'}$, the $r-\phi$ component of the shearing stress, which is typically assumed to be important for accretion disc. Following \cite{Chakrabarti1996}, we include the contribution to the ram pressure in the viscous stress in addition to the matter pressure, as given by $\sigma_\mathrm{r\phi}^{'}=\alpha(p+\rho v_\mathrm{r}^2)$. Note that throughout this work, the matter pressure, $p$, includes both gas and radiation. Here, $v_\mathrm{r}$, $v_\mathrm{\phi}$, $v_\mathrm{z}$ and $B_\mathrm{r}$, $B_\mathrm{\phi}$, $B_\mathrm{z}$ are respectively the $r$, $\phi$ and $z$ components of the velocity $\boldsymbol{v}$ and the magnetic field $\boldsymbol{B}$ respectively.

With all these assumptions, the equations (\ref{mass continuity}), (\ref{MHD equation}), (\ref{ion TdS}), (\ref{electron TdS}), (\ref{induction equation}), and (\ref{no monopole}) can be written as
\begin{equation}\label{continuity}
\frac{\partial}{\partial r}(r\rho v_\mathrm{r})=0,
\end{equation}
\begin{multline}\label{radial component NS}
v_\mathrm{r}\frac{\partial v_\mathrm{r}}{\partial r}-\frac{v^2 _\mathrm{\phi}}{r}=F_\mathrm{r}(r)-\frac{1}{\rho}\frac{\partial p}{\partial r}\\+\frac{1}{4\pi\rho}\left[\frac{-B^2 _\mathrm{\phi}}{r}+B_\mathrm{z}\frac{\partial B_\mathrm{r}}{\partial z}-B_\mathrm{\phi}\frac{\partial B_\mathrm{\phi}}{\partial r}-B_\mathrm{z}\frac{\partial B_\mathrm{z}}{\partial r}\right],
\end{multline}
\begin{multline}\label{azimuthal component NS}
	v_\mathrm{r}\frac{\partial v_\mathrm{\phi}}{\partial r}+\frac{v_\mathrm{r}v_\mathrm{\phi}}{r}=\frac{1}{\rho}\left[\frac{1}{r^2}\frac{\partial}{\partial r}(r^2 \sigma_\mathrm{r\phi}^{'})\right]\\+\frac{1}{4\pi\rho}\left[B_\mathrm{r}\frac{\partial B_\mathrm{\phi}}{\partial r}+\frac{B_\mathrm{r}B_\mathrm{\phi}}{r}+B_\mathrm{z}\frac{\partial B_\mathrm{\phi}}{\partial z}\right],
\end{multline}
\begin{equation}\label{energy equation for ion}
-v_\mathrm{r}T_\mathrm{i}\frac{dS}{dr}\bigg|_\mathrm{ion}=Q^\mathrm{+}-Q^\mathrm{ie},
\end{equation}
\begin{equation}\label{energy equation for electron}
-v_\mathrm{r}T_\mathrm{e}\frac{dS}{dr}\bigg|_\mathrm{electron}=Q^\mathrm{ie}-Q^\mathrm{rad},
\end{equation}
\begin{equation}\label{radial component induction}
\frac{\partial}{\partial z}(v_\mathrm{r}B_\mathrm{z})+\eta_\mathrm{B}\left[\frac{1}{r}\frac{\partial}{\partial r}\left(r\frac{\partial B_\mathrm{r}}{\partial r}\right)-\frac{B_\mathrm{r}}{r^2}+\frac{\partial^2B_\mathrm{r}}{\partial z^2}\right]=0,
\end{equation}
\begin{multline}\label{azimuthal component induction}
	\frac{\partial}{\partial z}(v_\mathrm{\phi} B_\mathrm{z})-\frac{\partial}{\partial r}(v_\mathrm{r} B_\mathrm{\phi}-v_\mathrm{\phi} B_\mathrm{r})\\+\eta_\mathrm{B}\left[\frac{1}{r}\frac{\partial}{\partial r}\left(r\frac{\partial B_\mathrm{\phi}}{\partial r}\right)-\frac{B_\mathrm{\phi}}{r^2}+\frac{\partial^2B_\mathrm{\phi}}{\partial z^2}\right]=0,
\end{multline}
\begin{equation}\label{vertical component induction}
-\frac{1}{r}\frac{\partial}{\partial r}(rv_\mathrm{r}B_\mathrm{z})+\eta_\mathrm{B}\left[\frac{1}{r}\frac{\partial}{\partial r}\left(r\frac{\partial B_\mathrm{z}}{\partial r}\right)+\frac{\partial^2B_\mathrm{z}}{\partial z^2}\right]=0,
\end{equation}
\begin{equation}\label{no monopole equation}
\frac{\partial B_\mathrm{r}}{\partial r}=-\frac{B_\mathrm{r}}{r}-\frac{\partial B_\mathrm{z}}{\partial z}.
\end{equation}
From the vertical equilibrium condition of the disc, due to contribution from gas and radiation pressures as well as magnetic field, the scale height of the disc can be written as,
\begin{equation}\label{scale height equation}
H(r)=r^{1/2}(r-2)\sqrt{\frac{\left(p+B_\mathrm{\phi}^2/8\pi\right)}{\rho}}.
\end{equation}
We assume that only the toroidal component of the magnetic field is vertically varying; that is why only the toroidal magnetic field appears in the vertical equilibrium equation. Also, in deriving this equation, the factor $\sqrt{r^2+z^2}$ present in the vertical component of force is approximated to $r$.

\subsection{Final equations}
\label{section final equations}
Finally, we reduce all the equations to the function of $r$ only by averaging vertically from 0 to $H$. To do that, we assume the vertical profiles for density and pressure as the conventional Keplerian profiles (\citealt{Pringle1981}), respectively, with
\begin{equation*}
	\rho=\rho_0exp(-z^2/(2H^2)),\ \text{and}\ p=p_0exp(-z^2/(2H^2)).
\end{equation*}
The quantities with subscript `0' represent the quantities at the mid-plane of the disc. The same vertical profiles for $\rho$ and $p$ indicate that the disc is vertically isothermal. We assume that the velocities $v_\mathrm{r}$, $v_\mathrm{\phi}$, and the electron temperature, $T_\mathrm{e}$, are independent of $z$. 

Various numerical simulations on the effects of LSMFs \cite[e.g.,][]{Machida2006, Oda2012} showed that the toroidal component of the magnetic field is able to hold the disc vertically. As $B_\mathrm{\phi}^2$ is equivalent to pressure, we assume
\begin{equation*}
	B^2_\mathrm{\phi}=B^2_\mathrm{\phi0}exp(-z^2/(2H^2)).
\end{equation*}
General relativistic magnetohydrodynamic simulation (\citealt{McKinney2012}) reported that $B_\mathrm{r}$ and $B_\mathrm{z}$ components grossly remain vertically uniform. We also assume the same, and that is why in scale height equation (\ref{scale height equation}) only $B_\mathrm{\phi}$ contributes. We put these vertical profiles in the equations (\ref{continuity}-\ref{no monopole equation}). When we integrate vertically all the terms from 0 to $H$, it brings terms containing $dH/dr$ as well as some numerical coefficients (i.e. $N_1$, $N_2$, ..). These numerical coefficients remain constant throughout the whole analysis. This leads to the vertically averaged eight coupled ordinary differential equations, which we solve to obtain the solution. The detailed final equations and the numerical coefficients are written in Appendix \ref{Appendix_final_equations}.

\subsection{Solution procedure}
\label{section solution procedure}
We solve a set of eight coupled ordinary differential equations, shown explicitly in Appendix in equations (\ref{averaged continuity}) - (\ref{averaged electron energy}) and (\ref{averaged azimuthal induction}) - (\ref{averaged no monopole}), using appropriate boundary conditions to obtain the solutions for eight flow variables: $v_\mathrm{r}$, $\lambda$, $B_\mathrm{r}$, $B_\mathrm{\phi}$, $B_\mathrm{z}$, $\rho$, $p$, and $T_\mathrm{e}$, as functions of the independent variable $r$. Here, $\lambda=r v_\mathrm{\phi}$ is the specific angular momentum of the flow. The outer boundary corresponds to the radius $r=r_\mathrm{out}$, at which $\lambda = \lambda_\mathrm{K}$, the Keplerian angular momentum per unit mass of the flow. It is basically the transition radius from the Keplerian to advective (sub-Keplerian) flows. The inner boundary corresponds to the event horizon of the black hole, at which the matter velocity reaches the light speed. Also, the black hole accretion is transonic in nature, i.e., the sub-sonic matter far away from the black hole passes through sonic/critical points as it drags inward, and becomes supersonic near the central black hole. We use such critical point location, $r=r_\mathrm{c}$, as one of the boundaries. Hence, the solutions connect the outer boundary to the black hole event horizon through critical/sonic location. 

To obtain the critical point condition, we combine the above mentioned equations (using Mathematica \citealt{Mathematica}) appropriately such that the slope of the radial velocity can be expressed in terms of all the flow variables and the independent variable $r$, as
\begin{equation*}
    \frac{dv_\mathrm{r}}{dr} = \frac{\mathcal{N}}{\mathcal{D}}.
\end{equation*}
The detailed expressions for $\mathcal{N}$ and $\mathcal{D}$ are given in equations (\ref{eq:numerator}) and (\ref{eq:denominator}), respectively. Following standard approach (\citealt{Chakrabarti1990}), the denominator of the slope vanishes at the critical radius $r_\mathrm{c}$. Existence of physical solution ensures that the numerator ($\mathcal{N}$) of the slope will also be zero at that radial point. Hence, at $r=r_\mathrm{c}$, $\mathcal{N} = \mathcal{D} =0$. As $dH/dr$ is involved in every equation, the following expression
\begin{equation}
	\frac{2}{H}\frac{dH}{dr}=\frac{1}{r}+\frac{2}{r-2}+\frac{(dp/dr)}{p+B_\mathrm{\phi}^2/8\pi}+\frac{2B_\mathrm{\phi}(dB_\mathrm{\phi}/dr)}{8\pi(p+B_\mathrm{\phi}^2/8\pi)}-\frac{1}{\rho}\frac{d\rho}{dr},
\end{equation}
is incorporated obtained from equation (\ref{scale height equation}). For optically thin flows, usually mass accretion rate ($\dot{M}$) lies within the range 10$^{-5}$-10$^{-2}$ $\dot{M}_\mathrm{Edd}$ ($\dot{M}_\mathrm{Edd}=L_\mathrm{Edd}/\eta c^2$, where $L_\mathrm{Edd}$ is the Eddington luminosity and we use efficiency factor $\eta=0.1$). Once we specify $\dot{M}$ of the system, from $\dot{M}=-4\pi N_1r\rho Hv_\mathrm{r}$, we can find the density ($\rho$) if we know the scale height ($H$) and radial velocity ($v_\mathrm{r}$) at that point.

Modeling of non-thermal electrons in radiatively inefficient accretion flows (\citealt{Yuan_et_al2003}) showed the range of electron temperature lying within $10^9-10^{10}$ K. \cite{Sarkar2020} used the entropy of the system to find the unique electron temperature for the transonic solution. However, the electron temperature is not a crucial parameter affecting our conclusion. That is why we fix the electron temperature at the critical point to $10^{10}$ K throughout our calculation. Following \cite{Mondal2018}, we use Alfven velocity in $\hat{r}$, $\hat{\phi}$ and $\hat{z}$ directions ($v_\mathrm{Arc}, v_\mathrm{A\phi c}, v_\mathrm{Azc}$) to fix the magnetic field at critical point. The subscript `c' denotes the values of the quantities at the critical point. The magnitude and sign of the magnetic field are fixed as follows,
\begin{equation}
v_\mathrm{Arc}=\frac{c_\mathrm{sc}}{\text{f}_\mathrm{rc}\sqrt{3}}, v_\mathrm{A\phi c}=\frac{c_\mathrm{sc}}{\text{f}_\mathrm{\phi c}\sqrt{3}}, v_\mathrm{Azc}=\frac{c_\mathrm{sc}}{\text{f}_\mathrm{zc}\sqrt{3}},\\
\end{equation}
which implies to
\begin{equation}
B_\mathrm{rc}=\sqrt{4\pi p/3}/\text{f}_\mathrm{rc}, B_\mathrm{\phi 0c}=\sqrt{4\pi p/3}/\text{f}_\mathrm{\phi c} \text{ and } B_\mathrm{zc}=\sqrt{4\pi p/3}/\text{f}_\mathrm{zc},\\
\end{equation}
where $c^2_\mathrm{sc} \sim p/\rho|_c$. Values of f$_\mathrm{ic}$ act as the inverse of the strength of the magnetic field. Here, $1/\sqrt{3}$ is the normalization factor for the three directions. We fix f$_\mathrm{rc}$, f$_\mathrm{\phi c}$, f$_\mathrm{zc}$ to fix the magnetic field at the critical point. These f$_\mathrm{ic}$ values fix only the magnetic field at the critical point. Radial evolution of the field is obtained from the solution of the eight coupled ordinary differential equations.

We fix the critical point location ($r_\mathrm{c}$) and the specific angular momentum ($\lambda_\mathrm{c}$) at $r_\mathrm{c}$ to their typical values. At the critical point, we simultaneously solve $\mathcal{N}=\mathcal{D}=0$, equation (\ref{scale height equation}) for scale height and the transcendental equation for the cut off frequency of synchrotron cooling (equation \ref{transcendental equation xM}) by using 4-dimensional Newton-Raphson root finding method. This solution simultaneously provides us $v_\mathrm{r}$, $H$, $\rho$, $p$, and subsequently all the physical parameters of the flow of our interest at the critical point.

Once values of all the parameters are obtained at the critical point, we proceed inward and outward from the critical point using the 4th order Runge-Kutta method to obtain complete solution. Assumption of steady state allows this procedure. However, to proceed from the critical point, we need to provide a slope of $v_\mathrm{r}$ at critical point, i.e., $(dv_\mathrm{r}/dr)_\mathrm{c}$. All other slopes can be represented in terms of $dv_\mathrm{r}/dr$. Traditionally $(dv_\mathrm{r}/dr)_\mathrm{c}$ used to be calculated using l'Hospital's rule (\citealt{Chakrabarti1996, Mukhopadhyay2015, Mondal2018}). The inclusion of magnetic field makes the equations much complex, and it becomes practically impossible to use l'Hospital's rule to find the exact value of $(dv_\mathrm{r}/dr)_\mathrm{c}$. That is why we use the trial and error method to find the approximate value of $(dv_\mathrm{r}/dr)_\mathrm{c}$. We first provide the approximate slope of $v_\mathrm{r}$ (all other slopes can be presented in terms of the slope of $v_\mathrm{r}$) at the critical point. Unless the slope value is close to the actual value, it does not give a smooth solution. By trial and error, we find the suitable value of the slope at the critical point. If the given slope is near to the actual value, within 1-2 radial steps, it converges to the actual value of the slope at that point. Proceeding outward, we stop at the outer boundary where $\lambda$ reaches $\lambda_\mathrm{K}$ if successful outward transport is possible and the disc forms. In the inner region, we stop when the total velocity reaches the speed of light, often before reaching the event horizon. As we use pseudo-Newtonian potential instead of the proper general relativistic framework, we face this limitation near the horizon of the black hole.

\section{Results}
\label{section Results}
By solving the eight coupled ordinary differential equations as described in section \ref{section assumptions over fundamental equations} (for details, see Appendix \ref{Appendix_final_equations}), we find the global solution of optically thin two temperature advective accretion disc in the presence of the LSMF. We explore the effect of the LSMF in the presence of turbulent viscosity in transporting angular momentum in stabilizing the advective accretion disc thermally at or above a critical accretion rate. Temperature ($T_\mathrm{i}$ and $T_\mathrm{e}$) and all the height integrated coolings are expressed in physical units, i.e. in K and ergs cm$^{-2}$ sec$^{-1}$ respectively. Magnetic fields are expressed in units of Gauss. Unless stated otherwise, all other variables are presented in standard dimensionless units, i.e., lengths are presented in units of gravitational radius ($r_\mathrm{g}$) where $r_\mathrm{g}=GM/c^2$, specific angular momentum ($\lambda$) is expressed in units of $GM/c$, velocities are presented in units of $c$, and $M$ is the mass of the black hole, which is fixed to $10 M_{\odot}$. We fix the following parameters at the critical point for all our investigations. Critical point location ($r_\mathrm{c}$) is fixed at 6.0 $r_\mathrm{g}$, specific angular momentum at critical point ($\lambda_\mathrm{c}$) is fixed at 3.0, electron temperature at critical point ($T_\mathrm{ec}$) is fixed at $10^{10}$K.

Initially, we assume a very weak vertical magnetic field and do all the analyses. We find that azimuthally dominated magnetic field (in comparison with radial field) with $B_\mathrm{\phi}$ directed in negative $\hat{\phi}$ direction while $B_\mathrm{r}$ is in positive $\hat{r}$ helps $\alpha$-viscosity the most in transporting angular momentum outward. On contrary, when $B_\mathrm{\phi}$ is directed in positive $\hat{\phi}$ direction and $B_\mathrm{r}$ is in negative $\hat{r}$ direction, the LSMF does the inward transport of angular momentum and opposes the effect of $\alpha$-viscosity. All these are presented in section \ref{section_result_suitable_magnetic_field_configuration} and we fix the magnetic field configuration depending on positive outward transportation of angular momentum for further investigations. The effect of the LSMF on disc dynamics as well as on thermal properties is described in section \ref{section mag effect} when it is helping the turbulent transport in the formation of the disc. After exploring the effect of the LSMF, we show that how with increasing accretion rate, the advection of energy tends to zero and then becomes negative. This leads the disc to be thermally unstable at or above the critical accretion rate and presented in section \ref{section instability due to mdot}. Finally, in the section \ref{section LSMF stabilizing disc} we show that thermally unstable optically thin discs can become stable with the help of the LSMF. A strong vertical field also can efficiently transport angular momentum outward like a strong toroidal field, which is shown in section \ref{section_vertical_field_transport}.

\subsection{Outward transport of angular momentum}
\label{section_result_suitable_magnetic_field_configuration}
\begin{figure}
\includegraphics[width=\columnwidth]{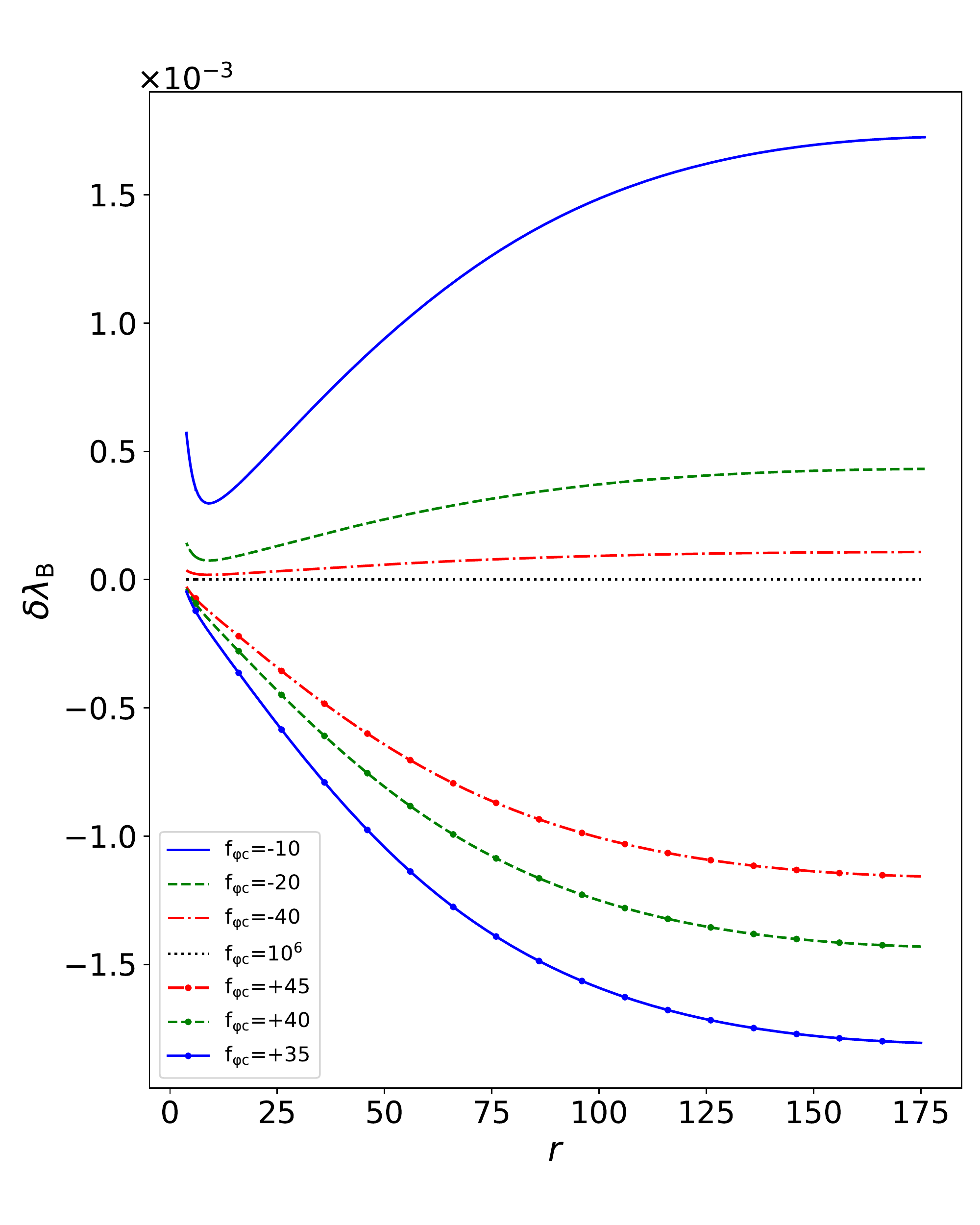}
\vspace{-0.5cm}
\caption{Magnetic contribution in transport of specific angular momentum ($\lambda$) for different magnitudes and orientation of $B_\mathrm{\phi}$. Positive value of $\delta\lambda_\mathrm{B}$ denotes the positive outward transport of angular momentum through the LSMF. $\alpha$ parameter is set to 0.02. The magnitude of magnetic field is set by f$_\mathrm{\phi c}$ value. Larger the f$_\mathrm{\phi c}$ value is, weaker the magnetic field is. $B_\mathrm{r}$ and $B_\mathrm{z}$ are in $\hat{r}$ and $\hat{z}$ direction always. $B_\mathrm{z}$ is kept very weak by fixing f$_\mathrm{zc}$=10$^6$. The magnitude of $B_\mathrm{r}$ is set to half of $B_\mathrm{\phi}$ at critical point. We fix $\dot{M}$=0.001 $\dot{M}_\mathrm{Edd}$.}
\label{fig mag helping opposing transport}
\end{figure}

Outward transport of angular momentum is essential for the formation of the disc. Initially, we do not consider the effect of vertical magnetic field and keep it negligible. Therefore, $B_\mathrm{r}$ and $B_\mathrm{\phi}$ are the main drivers for angular momentum transport through the LSMF. We find that depending on the magnitude and orientation of the magnetic field, the contribution of the LSMF in transporting angular momentum outward changes. This result is presented in Fig. \ref{fig mag helping opposing transport}. Shearing box simulations in the vertically stratified disc of angular momentum transport by MRI results in $\alpha=0.01-0.03$ without any net magnetic flux (\citealt{Davis2010, Simon2012}). That is why we choose $\alpha=0.02$ to represent the turbulent transport for our entire work. As this $\alpha$-viscosity is the result of turbulent transport only and the LSMF contributes separately for the transport of angular momentum, the effective value of $\alpha$ is calculated and justified with the observed value of $\alpha$ in section \ref{section Discussion}. Note that the observed value of $\alpha$ includes the contribution from all the sources of transport i.e., turbulent and LSMF for the present purpose.

To investigate the role of the magnetic field, we provide the magnetic field with different orientations and magnitudes at the critical point. This is the $\boldsymbol{B}$ at critical point for our calculation, depending on which the magnetic field configuration is evolved over the radius. We can write the equation of transport of specific angular momentum (\ref{final equation angular momentum transport} in Appendix) in the following form:
\begin{equation}
\label{equation_dlambda/dr}
\frac{d\lambda}{dr}=\Big(\frac{d\lambda}{dr}\Big)_\mathrm{\alpha}+\Big(\frac{d\lambda}{dr}\Big)_\mathrm{B}=\delta \lambda_\mathrm{\alpha}+\delta \lambda_\mathrm{B},
\end{equation}
where $\delta \lambda_\mathrm{\alpha}$ and  $\delta \lambda_\mathrm{B}$ are the transports based on $\alpha$ and LSMF contributions respectively in $d\lambda/dr$. By rearranging the terms we can write from equation (\ref{final equation angular momentum transport})
\begin{multline}
\label{equation_dlambda/dr_B}
\delta \lambda_\mathrm{B}=\frac{r}{4\pi\rho_0v_\mathrm{r}}\Big[N_4\Big(B_\mathrm{r}\frac{dB_\mathrm{\phi 0}}{dr}+\frac{B_\mathrm{r}B_\mathrm{\phi 0}}{r}\Big)\\+N_5B_\mathrm{r}B_\mathrm{\phi 0}\Big(\frac{1}{H}\frac{dH}{dr}\Big)+\frac{N_6}{H}B_\mathrm{z}B_\mathrm{\phi 0}\Big].
\end{multline} 
Here, the quantities which vary vertically, are represented with subscript `0' at mid-plane. Vertical integration of flow variables leads to different numerical coefficients, i.e., $N_4$, $N_5$, $N_6$ (for details see Appendix \ref{Appendix_final_equations}). These numerical coefficients remain constant throughout our whole analysis. The positive value of $d\lambda/dr$ denotes the successful outward transport of angular momentum. For positive value of $\alpha$, $\delta \lambda_\mathrm{\alpha}$ is always positive. Now, depending on whether $\delta \lambda_\mathrm{B}$ is positive or negative, the LSMF supports or opposes $\alpha$-viscosity in transporting angular momentum outward. The positive value of $\delta \lambda_\mathrm{B}$ denotes the successful outward transport of angular momentum through the LSMF. $\delta \lambda_\mathrm{B}$ has two contributions: from that involved with $B_\mathrm{r}$ and $B_\mathrm{\phi}$, and from that involved with $B_\mathrm{z}$ and $B_\mathrm{\phi}$. Although contribution through $B_\mathrm{r}$ and $B_\mathrm{\phi}$ involves term $dB_\mathrm{\phi}/dr$ which makes the situation complex in determining the orientation and magnitude of $B_\mathrm{r}$ and $B_\mathrm{\phi}$ for the positive value of $\delta \lambda_\mathrm{B}$, the product $B_\mathrm{r}B_\mathrm{\phi}$ remains the dominant component. After considering negative sign of $v_\mathrm{r}$ and $N_6$, we find that positive value of $\delta \lambda_\mathrm{B}$ is possible when $B_\mathrm{r}B_\mathrm{\phi}$ is negative and $B_\mathrm{z}B_\mathrm{\phi}$ is positive. As we are solving equations in upper half-plane of the disc, $B_\mathrm{r}$ and $B_\mathrm{z}$ can be fixed in positive $\hat{r}$ and positive $\hat{z}$ (or in negative $\hat{r}$ and negative $\hat{z}$) directions. All the above constraints together indicate that depending on the direction of $B_\mathrm{\phi}$, any one component, either $B_\mathrm{r}B_\mathrm{\phi}$ or $B_\mathrm{z}B_\mathrm{\phi}$, will contribute positively in $\delta \lambda_\mathrm{B}$ and the other component will oppose. For the time being, we ignore any effect of the vertical magnetic field by keeping $B_\mathrm{z}$ very weak and focusing on angular momentum transport through $B_\mathrm{r}$ and $B_\mathrm{\phi}$ along with $\alpha$-viscosity. To make $B_\mathrm{r}B_\mathrm{\phi}$ negative, $B_\mathrm{\phi}$ has to be in negative $\hat{\phi}$ direction. With this configuration, weak $B_\mathrm{z}$ also keeps the opposite contribution of $B_\mathrm{z}B_\mathrm{\phi}$ small. In addition, we find that if we make $B_\mathrm{\phi 0c}$ equal or weaker compared to $B_\mathrm{rc}$, then $B_\mathrm{\phi}$ changes direction, and $B_\mathrm{r}B_\mathrm{\phi}$ contributes oppositely in outward transport of angular momentum before reaching $\lambda=\lambda_\mathrm{K}$, and strong enough field with this configuration can even nullify the positive contribution of $\alpha$-viscosity and disc may not form. To keep things simple, we restrict our parameter space such that $B_\mathrm{\phi}$ remains unidirectional throughout the whole radial range. Also, $B_\mathrm{\phi}$ can not be too strong compared to $B_\mathrm{r}$. Otherwise, the opposite contribution from $B_\mathrm{z}B_\mathrm{\phi}$ will start to give effect although $B_\mathrm{z}$ is small. In this way, we find that the most efficient, positive transport due to contribution from $B_\mathrm{r}$ and $B_\mathrm{\phi}$ occurs when the magnitude of the toroidal magnetic field is twice of the radial magnetic field at the critical point, i.e., $B_\mathrm{\phi 0c}=2 B_\mathrm{rc}$. 

Fig. \ref{fig mag helping opposing transport} shows the positive and negative contributions of the LSMF in outward transport of angular momentum depending on the direction of $B_\mathrm{\phi}$. $B_\mathrm{z}$ is kept negligible by fixing f$_{zc}=10^6$, and the magnitude of $B_\mathrm{\phi 0c}$ is always kept two times of $B_\mathrm{rc}$ as described above. The directions of $B_\mathrm{r}$ and $B_\mathrm{z}$ are kept fixed in positive $\hat{r}$ and positive $\hat{z}$ directions whereas $B_\mathrm{\phi}$ changes direction between negative $\hat{\phi}$ and positive $\hat{\phi}$ which is indicated by negative and positive values of f$_\mathrm{\phi c}$ respectively. The top blue solid line indicates the contribution of the LSMF in outward transport of angular momentum when $B_\mathrm{\phi}$ is strong and in negative $\hat{\phi}$ direction, which is indicated by f$_\mathrm{\phi c}=-10$. Gradually as the field becomes weaker (magnitude of f$_\mathrm{\phi c}$ increases), the magnetic contribution decreases. When f$_\mathrm{\phi c}=10^6$, there is practically zero magnetic field and zero contribution in transport as indicated by the black dotted line. This is the situation when the whole transport is governed by turbulent viscosity only, which is approximated through $\alpha=0.02$ for all our calculations. Now, as $B_\mathrm{\phi}$ reverses direction to positive $\hat{\phi}$, $\delta \lambda_\mathrm{B}$ becomes negative and does inward transport of angular momentum instead of outward. Also, in Fig. \ref{fig mag helping opposing transport}, the positive values of f$_\mathrm{\phi c}$ are relatively higher than negative values, although the magnitude of $\delta \lambda_\mathrm{B}$ remains similar. Hence, we can conclude that for toroidally dominated LSMF, to make angular momentum transport outward, $B_\mathrm{\phi}$ should be in negative $\hat{\phi}$ direction, and, to make the angular momentum transport inward of similar magnitude, relatively weaker $B_\mathrm{\phi}$ in positive $\hat{\phi}$ direction is required. The above scenario indicates that with a weak vertical field when $B_\mathrm{r}$ and $B_\mathrm{z}$ are in positive $\hat{r}$ and positive $\hat{z}$ directions respectively and relatively strong $B_\mathrm{\phi}$ is in negative $\hat{\phi}$ direction, the LSMF can efficiently transport angular momentum outward and helps turbulent $\alpha$-viscosity in the formation of the disc. As outward transport of angular momentum is necessary for the formation of the disc, we fix the magnetic field with this configuration for our investigations. In section \ref{section_vertical_field_transport} we show that a strong vertical magnetic field with the help of a toroidal field in a suitable direction can also transport the angular momentum efficiently. 

\begin{figure}
\includegraphics[width=\columnwidth]{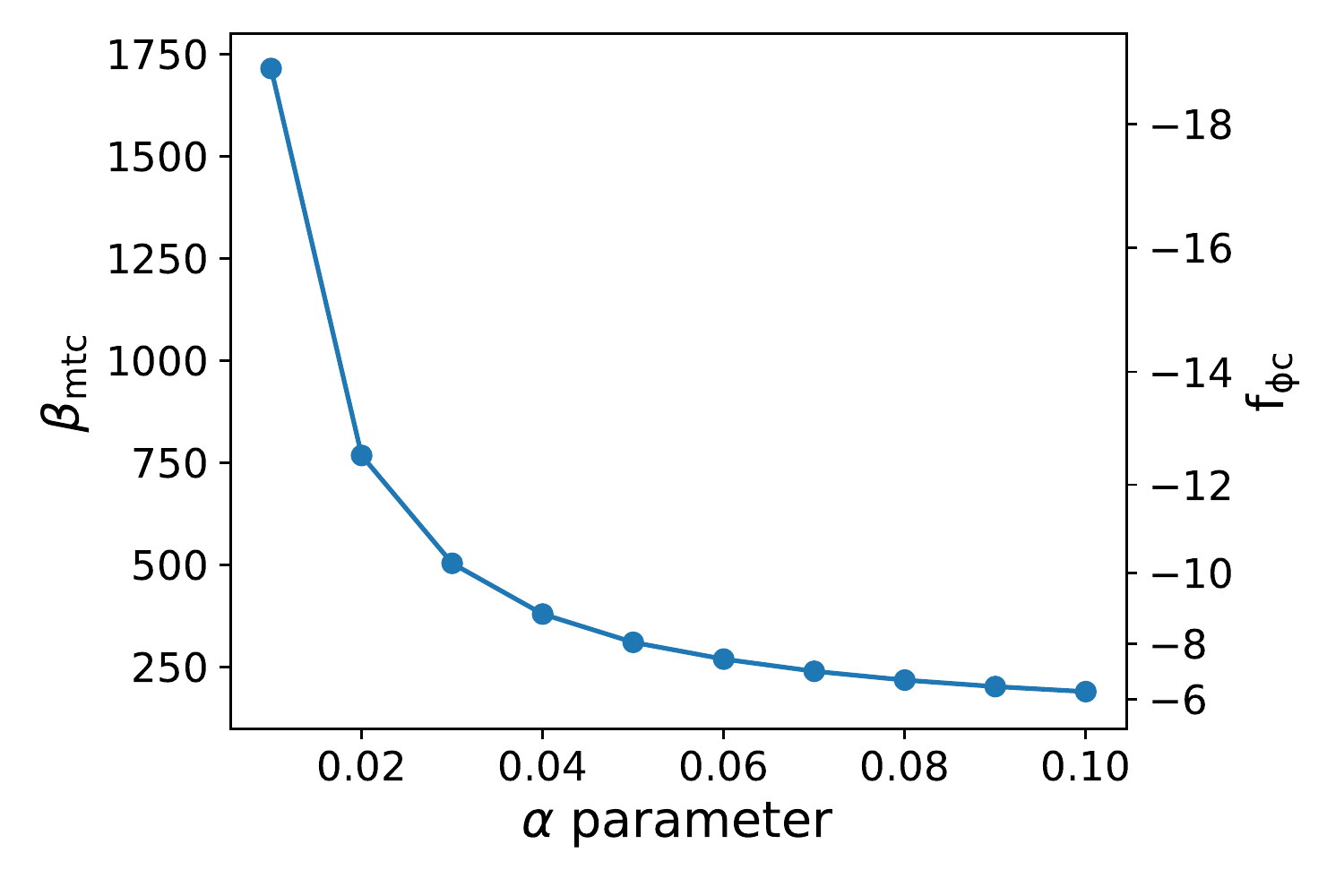}
\caption{The threshold $\beta_\mathrm{m}$ at critical point, $\beta_\mathrm{mtc}$, with the change of $\alpha$ parameter. Larger the $\beta_\mathrm{mtc}$ is, weaker the magnetic field is. Corresponding f$_\mathrm{\phi c}$ value is shown in the right hand vertical axis. Magnitude of $B_\mathrm{\phi 0c}$ is twice of $B_\mathrm{rc}$ for all the calculations while $B_\mathrm{z}$ is negligible. $\dot{M}$ is fixed at 0.001 $\dot{M}_\mathrm{Edd}$.}
\label{threshold magnetic field}
\end{figure}

Before finding the effect of the LSMF on dynamics and its thermal properties, it is important to find the threshold value of the LSMF above which it starts to affect the properties of the disc. Other than the LSMF, the disc's properties are governed by the value of $\alpha$-parameter. That is why it is expected that threshold LSMF will change with the $\alpha$ value, i.e., with the turbulent transport parameter of the disc. To define the threshold, we first integrate equation \eqref{equation_dlambda/dr} numerically over the radial range of the disc and find  
\begin{equation*}
	\Delta\lambda=\Delta\lambda_\mathrm{\alpha}+\Delta\lambda_\mathrm{B}.
\end{equation*}
$\Delta\lambda$ gives the total transported angular momentum outward for the whole disc. $\Delta\lambda_\mathrm{\alpha}$ and $\Delta\lambda_\mathrm{B}$ measure the $\alpha$-viscosity and LSMF contributions in transport respectively for the whole disc. We define the threshold magnetic field value at the critical point ($B_\mathrm{tc}$) when $\Delta \lambda_\mathrm{B}$ becomes one percentage of $\Delta \lambda$, i.e., LSMF contribution becomes one percent of total transported angular momentum. With increasing the value of $\alpha$-parameter naturally the outward transportation of angular momentum through turbulent $\alpha$-viscosity increases. Consequently, to make the effect of the LSMF in outward transportation significant over $\alpha$-viscosity, $B_\mathrm{tc}$ has to increase and correspondingly $\beta_\mathrm{mtc}$ has to decrease. This result is shown in Fig. \ref{threshold magnetic field}. Value of f$_\mathrm{\phi c}$ is presented in the right side vertical axis. We can see that for f$_\mathrm{\phi c}$ values around -10, $\beta_\mathrm{mtc}$ $\sim$ 500. It is quite encouraging that the value of $\beta_\mathrm{mtc}$ is quite large even for $\alpha=0.1$. Hence, weak LSMF starts to contribute to transportation and can affect the disc dynamics, which was earlier shown by \cite{Mukhopadhyay2015}.

\subsection{Effect of large-scale magnetic field}
\label{section mag effect}

The evolution of the magnetic field and its effect on the disc is quite complex as it involves many coupled equations. The main goal of this work is to show that the LSMF can transport angular momentum and thermally stabilize the disc. Thermal stabilization of the disc depends on the advection factor of ion, which is solely governed by heating and cooling of ion. As long as the advected energy through ion is positive, the optically thin disc is thermally stable. Depending on the necessary requirement of outward transport of angular momentum through the LSMF, we have already fixed the orientation and relative magnitude of $B_\mathrm{\phi}$ and $B_\mathrm{r}$ at the critical point with keeping $B_\mathrm{z}$ negligible. In this section, in addition to the positive transport of angular momentum, we present the effect of the LSMF on disc dynamics and the heating and cooling of ions and electrons.

\subsubsection{Effects on dynamics}
\label{section mag effect dynamics}
\begin{figure*}
\centering\includegraphics[scale=0.58]{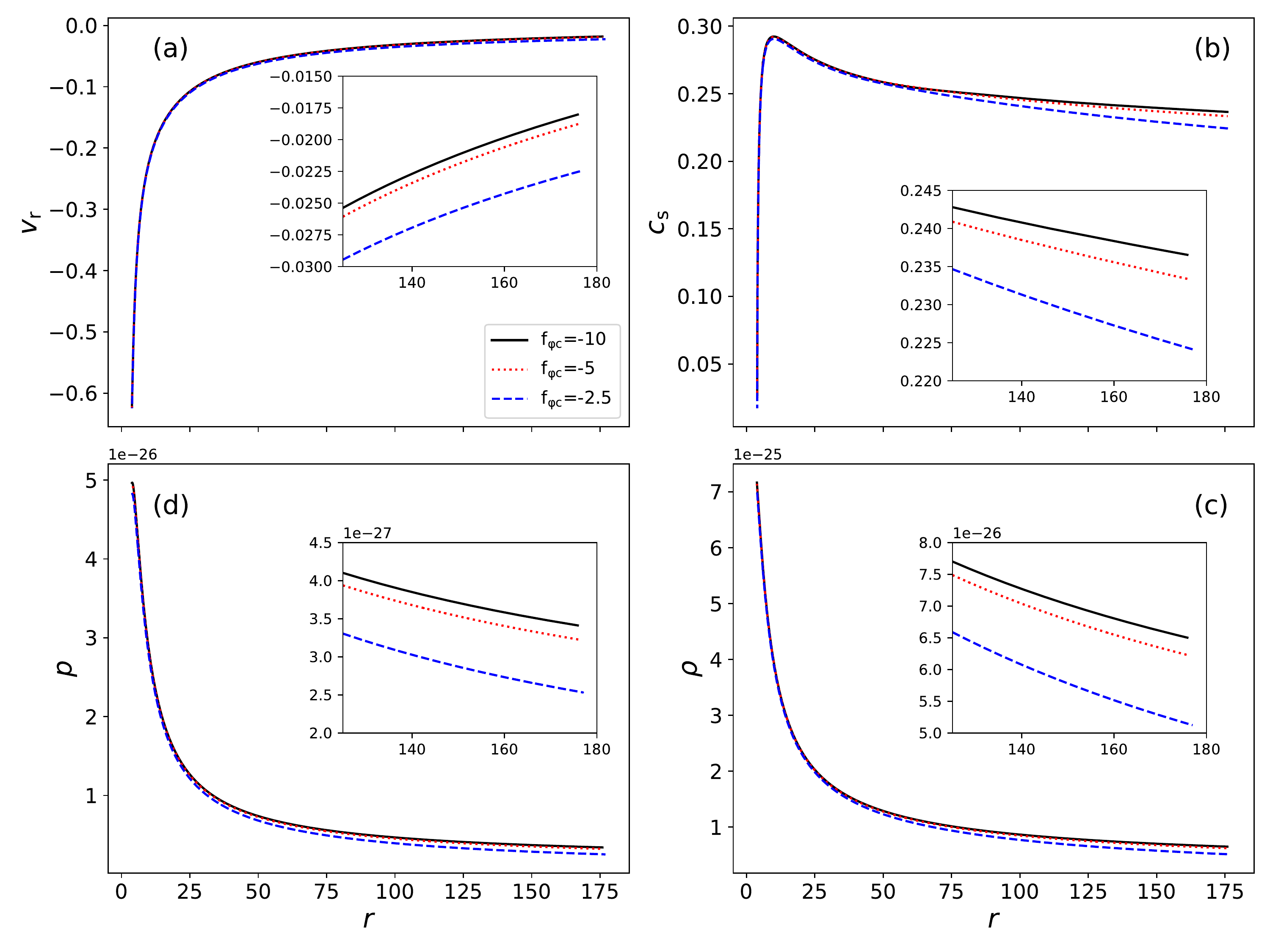}
\caption{Variation of radial velocity, sound speed, density and pressure with the magnetic field strength. Zoomed in view for the outer region of the disc is shown for each variable to show the effect of magnitude of magnetic field more clearly. Magnitude of $B_\mathrm{\phi 0c}$ is twice of $B_\mathrm{rc}$ along with negligible $B_\mathrm{z}$ for all the calculations. $\alpha$=0.02 and $\dot{M}$=0.001 $\dot{M}_\mathrm{Edd}$ are fixed.}
\label{fig_mag_effect_vr_cs}
\end{figure*}

In Fig. \ref{fig_mag_effect_vr_cs} we present the radial evolution of $v_\mathrm{r}$, $c_\mathrm{s}$, $\rho$ and $p$ for different strengths of magnetic field. The lower the magnitude of f$_\mathrm{\phi c}$, the stronger the magnetic field is. Keep in mind that magnetic field configuration is fixed so that it transports angular momentum outward as like $\alpha$-viscosity and helps turbulent viscosity for the formation of the disc. Zoomed in view for the outer region of the disc is shown for each variable to show the effect of the magnitude of magnetic field more clearly. In Fig. \ref{fig_mag_effect_vr_cs}(a), $v_\mathrm{r}$ is plotted. Negative values of $v_\mathrm{r}$ indicate accretion as matter flows in negative $\hat{r}$ direction. With increasing magnetic field strength, the magnitude of $v_\mathrm{r}$ also increases, i.e., the matter is advected more rapidly. With increasing magnetic field strength as radial inward velocity increases (negative value of $v_\mathrm{r}$), it reduces $\rho$ due to constancy of $\dot{M}$. Pressure also decreases with increasing magnetic field strength. This pressure includes gas (ion and electron) and radiation. As we consider radiatively inefficient optically thin flow, ions give the largest contribution in pressure. Later we show that the work done by the plasma due to compression or expansion plays a crucial role in decreasing ion temperature and subsequently the ion pressure with the increasing strength of the magnetic field. This is further explained in section \ref{section mag effect thermal}. The decrement in pressure is larger than the decrement in density, leading to the decrement in sound speed with the increasing strength of the magnetic field. This is shown in Fig. \ref{fig_mag_effect_vr_cs}(b).

\begin{figure*}
\includegraphics[scale=0.39]{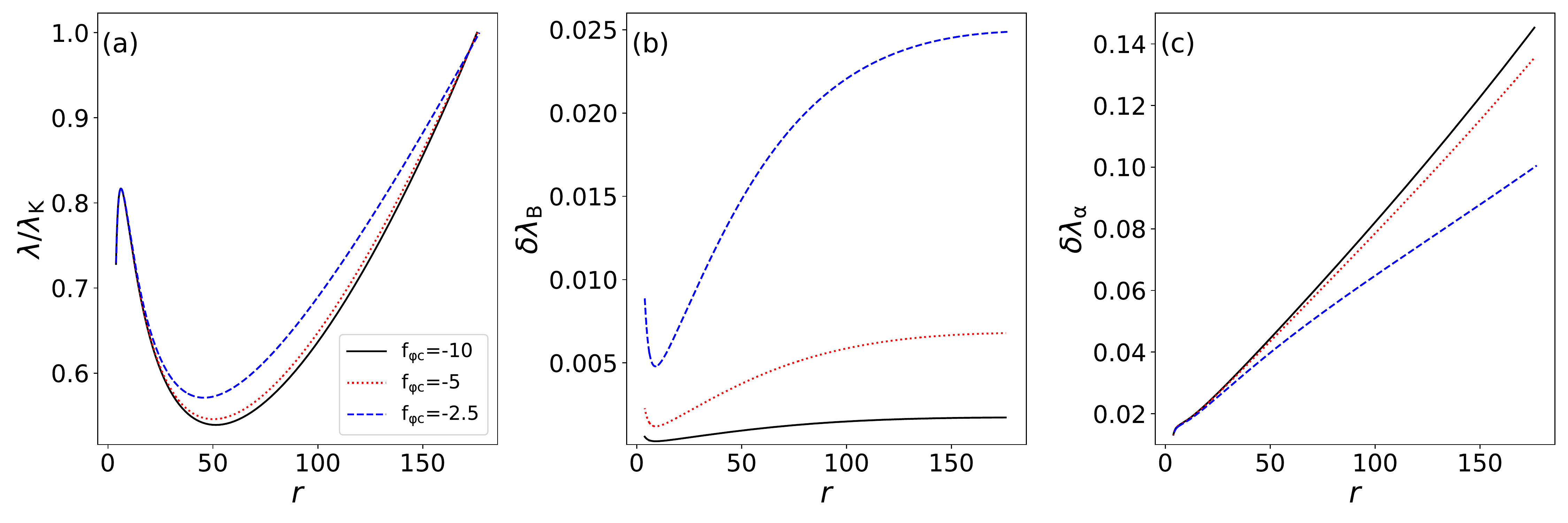}
\caption{Variation of various quantities related to angular momentum transport with magnetic field strength. Magnetic field configuration and all other parameters are set as in Fig. \ref{fig_mag_effect_vr_cs}.}
\label{fig_mag_effect_lambda}
\end{figure*}

In Fig. \ref{fig_mag_effect_lambda} we present how outward transport of angular momentum changes with the strengths of the magnetic field. $\lambda/\lambda_\mathrm{K}$ is plotted in Fig. \ref{fig_mag_effect_lambda}(a) and correspondingly the contributions from magnetic field and $\alpha$-viscosity in transportation i.e. $\delta\lambda_\mathrm{B}$ and $\delta\lambda_\mathrm{\alpha}$ are plotted in Fig. \ref{fig_mag_effect_lambda}(b) \& \ref{fig_mag_effect_lambda}(c). Although, as per our expectation, $\delta\lambda_\mathrm{B}$ increases with the increasing strength of the field, $\delta\lambda_\mathrm{\alpha}$ decreases, i.e., turbulent transport becomes weak in transporting angular momentum outward with the stronger magnetic field. The ratio of pressure and density ($\alpha p_0/\rho_0$ term in equation \eqref{final equation angular momentum transport}), i.e., the temperature of the flow has the dominant effect in outward transport of angular momentum through $\alpha$-viscosity. This means $\alpha$-viscosity transports angular momentum more efficiently for the hotter disc. Thus stronger magnetic field reduces the turbulent transport by making it cooler. The decrement in turbulent transport, $\delta\lambda_\mathrm{\alpha}$, is almost balanced by the increment in $\delta\lambda_\mathrm{B}$, and the evolution of $\lambda/\lambda_\mathrm{K}$, i.e., the evolution of angular momentum remains almost similar for different strengths of the magnetic field. The outer boundary condition for the truncation of the disc is $\lambda=\lambda_\mathrm{K}$. That is why disc size remains almost similar for different magnetic field strengths. Changing magnetic field strength changes the relative contributions from the $\alpha$-viscosity and the LSMF in outward transportation, keeping the overall transport the same. Even for f$_\mathrm{\phi c}=-2.5$, $\delta\lambda_\mathrm{B}$ remains smaller than $\delta\lambda_\mathrm{\alpha}$. This indicates that even for the strongest field value, the main driver for the transport remains turbulent viscosity which is approximated through $\alpha=0.02$. However, the assumption of constant $\alpha$ value may not be the case in reality.

\begin{figure*}
\includegraphics[scale=0.39]{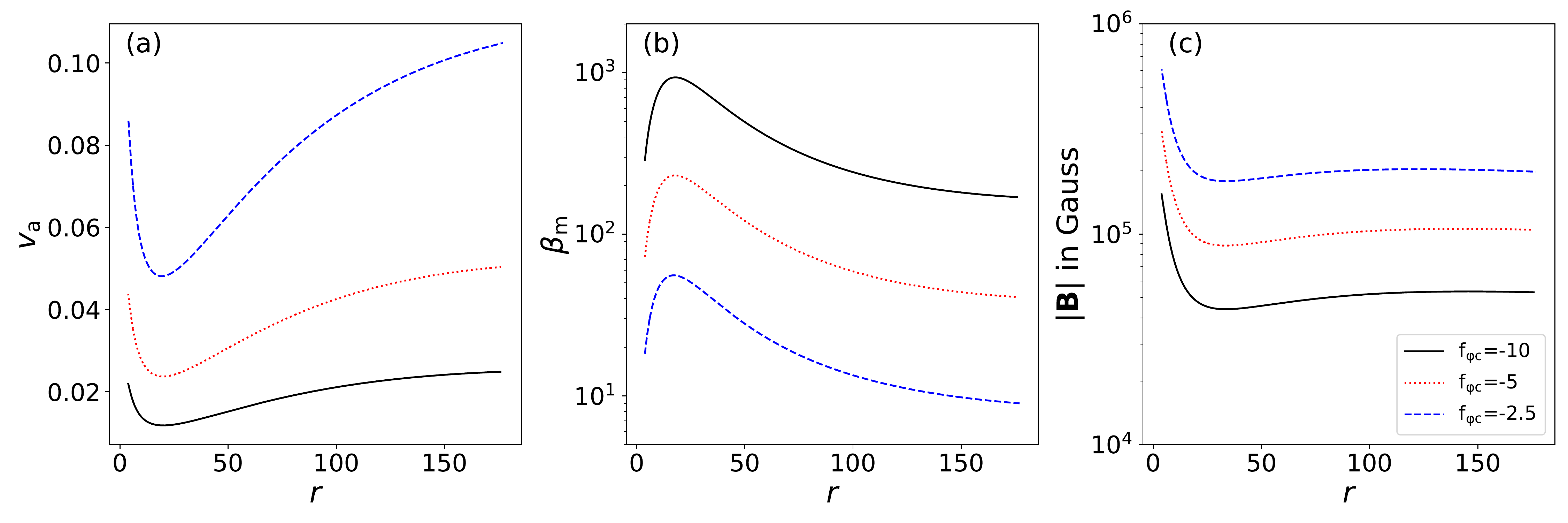}
\caption{Variation of Alfven velocity, $\beta_\mathrm{m}$ and magnitude of magnetic field with the strength of magnetic field. Magnetic field configuration and all other parameters are set as in Fig. \ref{fig_mag_effect_vr_cs}.}
\label{fig_mag_effect_Alfven_betam}
\end{figure*}

In Fig. \ref{fig_mag_effect_Alfven_betam} we present the evolution of Alfven velocity ($v_a$), $\beta_\mathrm{m}$ and the magnitude of magnetic field (|$\boldsymbol{B}$|) which explicitly denote the relative dominance of magnetic field in the disc. Naturally, with increasing strength of the field, $v_a$ increases, and $\beta_\mathrm{m}$ decreases. Radially |$\boldsymbol{B}$| remains almost constant in the outer region of the disc and increases in the inner region. This indicates that in the disc's outer region, the magnetic field's inward advection is balanced by its diffusion; high inward radial advection in the inner region dominates over diffusion and increases the magnitude of the field. However, the pressure (ion+electron+radiation) and density decrease in the outer region. That is why although |$\boldsymbol{B}$| remains almost constant, $v_a$ increases, and $\beta_\mathrm{m}$ decreases in the outer region of the disc. Within our investigated parameter space, even for the strongest field strength, the value of $\beta_\mathrm{m}$ remains within the range of 5-60 over the whole radial range of the disc. If we assume that the source of turbulent viscosity, i.e., $\alpha$-viscosity is MRI, then $\beta_\mathrm{m}\gtrsim 5$ confirms that MRI remains active and there is no restriction in applying the LSMF along with the $\alpha$-viscosity. Even the value of $\beta_\mathrm{m}$ is not restricted for MRI to be active if the disc size is not limited to thin (\citealt{Kim2000}).

\subsubsection{Effects on heating and cooling}
\label{section mag effect thermal}
\begin{figure}
\includegraphics[width=\columnwidth]{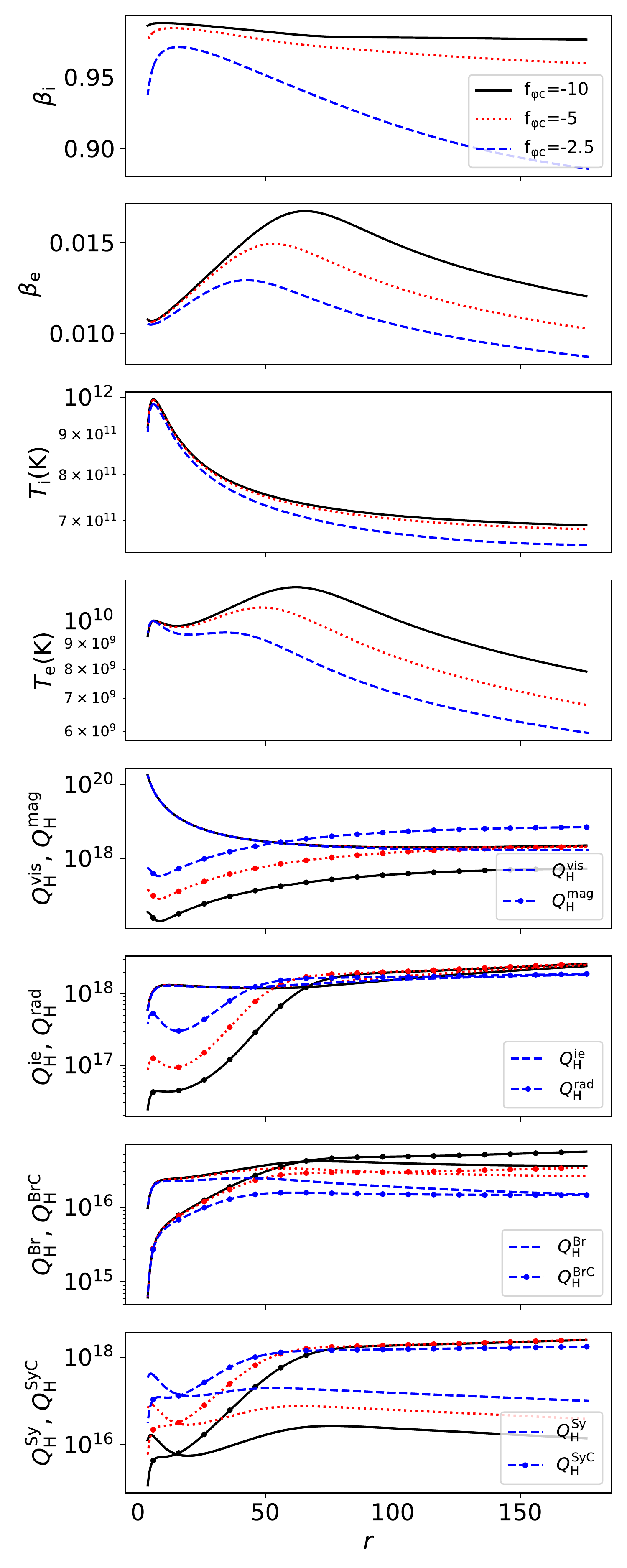}
\caption{Variation of various variables related to thermal properties of the disc with magnetic field strength. All height integrated heating and cooling are in units of ergs cm$^{-2}$ sec$^{-1}$. Magnetic field configuration as well as other parameters are same as in Fig.\ref{fig_mag_effect_vr_cs}.}
\label{mag effect thermal variables}
\end{figure}

Once the dynamical variables are affected by the LSMF, it is evident that the disc's heating and cooling will also be affected. This subsection discusses how temperatures of ions and electrons and how different heating and cooling mechanisms vary with the strength of the magnetic field. The radial variation of corresponding physical variables with the magnetic field is plotted in Fig. \ref{mag effect thermal variables}. 

For ions, viscous dissipation ($Q^\mathrm{vis}_\mathrm{H}$,  $H$ is written in subscript as the quantity is vertically integrated over the scale height $H$) and Joule heating ($Q^\mathrm{mag}_\mathrm{H}$) are the sources of heating, whereas heat transfers to electrons through ion-electron Coulomb coupling ($Q^\mathrm{ie}_\mathrm{H}$) cools the ion. On the other hand, $Q^\mathrm{ie}_\mathrm{H}$ heats the electrons and different radiation mechanisms (bremsstrahlung ($Q^\mathrm{Br}_\mathrm{H}$), synchrotron ($Q^\mathrm{Sy}_\mathrm{H}$) and their comptonizations ($Q^\mathrm{BrC}_\mathrm{H}$, $Q^\mathrm{SyC}_\mathrm{H}$)) together, $Q^\mathrm{rad}_\mathrm{H}$ serves as cooling mechanism. Magnetic field explicitly affects the Joule heating ($Q^\mathrm{mag}_\mathrm{H}$) as well as the synchrotron cooling ($Q^\mathrm{Sy}_\mathrm{H}$) and its comptonization ($Q^\mathrm{SyC}_\mathrm{H}$). The effect on $Q^\mathrm{vis}_\mathrm{H}$, $Q^\mathrm{ie}_\mathrm{H}$, and on other cooling mechanisms, $Q^\mathrm{Br}_\mathrm{H}$, $Q^\mathrm{BrC}_\mathrm{H}$ comes through the change in density and temperature of ions and electrons. 

Although $Q^\mathrm{vis}_\mathrm{H}$ is hardly affected, $Q^\mathrm{mag}_\mathrm{H}$ increases naturally with the increasing strength of the magnetic field. Even $Q^\mathrm{mag}_\mathrm{H}$ dominates over $Q^\mathrm{vis}_\mathrm{H}$ in the outer region of the disc for f$_\mathrm{\phi c}=-2.5$ and acts as the primary heating for the ions. Therefore, heating for ions increases with the increasing strength of the magnetic field.

$Q^\mathrm{Sy}_\mathrm{H}$ and $Q^\mathrm{SyC}_\mathrm{H}$ both increase significantly with increasing magnetic field strength. However, as the stronger field reduces the density, consequently $Q^\mathrm{Br}_\mathrm{H}$ and its comptonization decrease with the stronger magnetic field. For our investigated parameter space, the magnitude of Synchrotron cooling is more than the bremsstrahlung, and as a net effect, the total cooling ($Q^\mathrm{rad}_\mathrm{H}$) increases with the strength of the magnetic field. However, $Q^\mathrm{ie}_\mathrm{H}$ remains almost constant for different strengths of the magnetic field due to the complex dependence on density and on the difference between ion and electron temperature. 

$\beta_\mathrm{i}$ and $\beta_\mathrm{e}$ represent ion and electron fractions respectively of total pressure which includes ion, electron, radiation and magnetic field (equation \ref{definition beta}). We find that with the increasing strength of the magnetic field, $\beta_\mathrm{i}$ and $\beta_\mathrm{e}$ decrease. The reason is twofold. With a stronger magnetic field, the radiation and magnetic contribution in total pressure naturally increase and reduce the ion and electron fraction in total pressure. In addition to that, the ion and electron temperature decrease with the increasing strength of the magnetic field. For electron, it is quite easy to understand. Electrons' heating ($Q^\mathrm{ie}_\mathrm{H}$) remains almost constant whereas its cooling ($Q^\mathrm{rad}_\mathrm{H}$) increases with increasing strength of magnetic field. This naturally will cool down the electrons, leading to the decrement in temperature with the stronger magnetic field. However, for ions, with increasing strength of the magnetic field, even though its net heating ($Q^\mathrm{vis}_\mathrm{H}+Q^\mathrm{mag}_\mathrm{H}$-$Q^\mathrm{ie}_\mathrm{H}$), i.e., the advected entropy increases, still they cool down. Here comes an interesting thing. With one step further, the temperature evolution does not depend only on the net heating; it depends on how much the increment in net heating contributes to the increment in internal energy and the increment in work done by the plasma.

\begin{figure}
\includegraphics[width=\columnwidth]{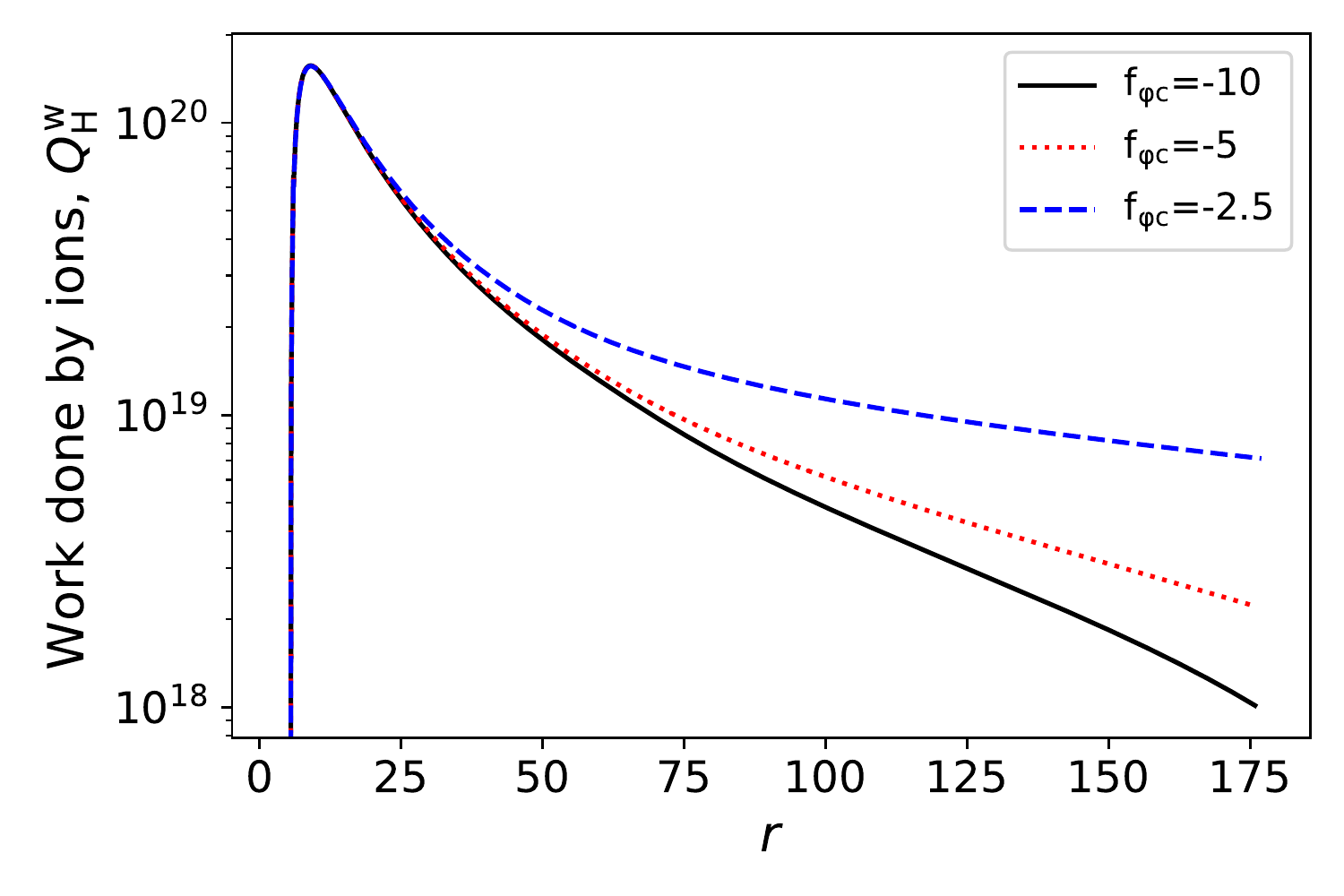}
\caption{Variation of work done by the gas due to compression or expansion with the different strengths of the magnetic field. The height integrated work done by the gas is presented in units of ergs cm$^{-2}$ sec$^{-1}$. Magnetic field configuration as well as other parameters are same as in Fig.\ref{fig_mag_effect_vr_cs}. With the increasing strength of the magnetic field, the ions do larger amount of work which is leading to the decrement of temperature with the increasing strength of magnetic field.}
\label{fig_compression_work}
\end{figure}

From the first law of thermodynamics, we know that the heat energy ($dQ$) supplied to a system is used only partially to increase its internal energy ($dU$), with the rest of the heat energy, the system does some work ($dW$), $dQ=dU+dW$. Now, if the work done by the system ($dW$) exceeds the supplied heat energy ($dQ$), then it does the work at the expense of its internal energy and will cool down the system. This is the case that is cooling down the ions here even if the heat energy supplied to ions increases with the increment of the strength of the magnetic field. We have found that with the increment of the strength of the magnetic field, the work done by the ions (work done due to compression or expansion, $Q^\mathrm{w}_\mathrm{H}$) increases more than the supplied heat energy ($Q^\mathrm{vis}_\mathrm{H}+Q^\mathrm{mag}_\mathrm{H}$). The expression for the height integrated work done by the ions is given by
\begin{equation*}
Q^\mathrm{w}_\mathrm{H}=\int_{0}^{H}(-v_\mathrm{r})\rho p_\mathrm{i}\frac{dV}{dr}=(-v_\mathrm{r})(-p_\mathrm{i0})(\frac{N_1H}{\rho_0}\frac{d\rho_0}{dr}+\frac{N_2}{2}\frac{dH}{dr}),
\end{equation*}
where $p_\mathrm{i}$ is the ion pressure, $V$ is the volume, $p_\mathrm{i0}$ is the ion pressure at the mid-plane. In deriving this expression we have assumed the vertical variation of $p_\mathrm{i}$ is same as $p$. This approximation is justifiable as ion pressure gives the main contribution in total pressure. This is evident from the value of $\beta_\mathrm{i}$ which remains around 0.95 for this parameter regime. In Fig. \ref{fig_compression_work} we show the variation of work done by the ions for different strengths of the magnetic field. This height integrated work done is plotted in units of ergs cm$^{-2}$ sec$^{-1}$. In summary, the increment in $dW$ dominates over $dQ$ which finally is leading to the decrement in the internal energy and finally makes the ions cooler.

\subsection{Effect of higher accretion rate and thermal instability}
\label{section instability due to mdot}
\begin{figure}
\includegraphics[width=\columnwidth]{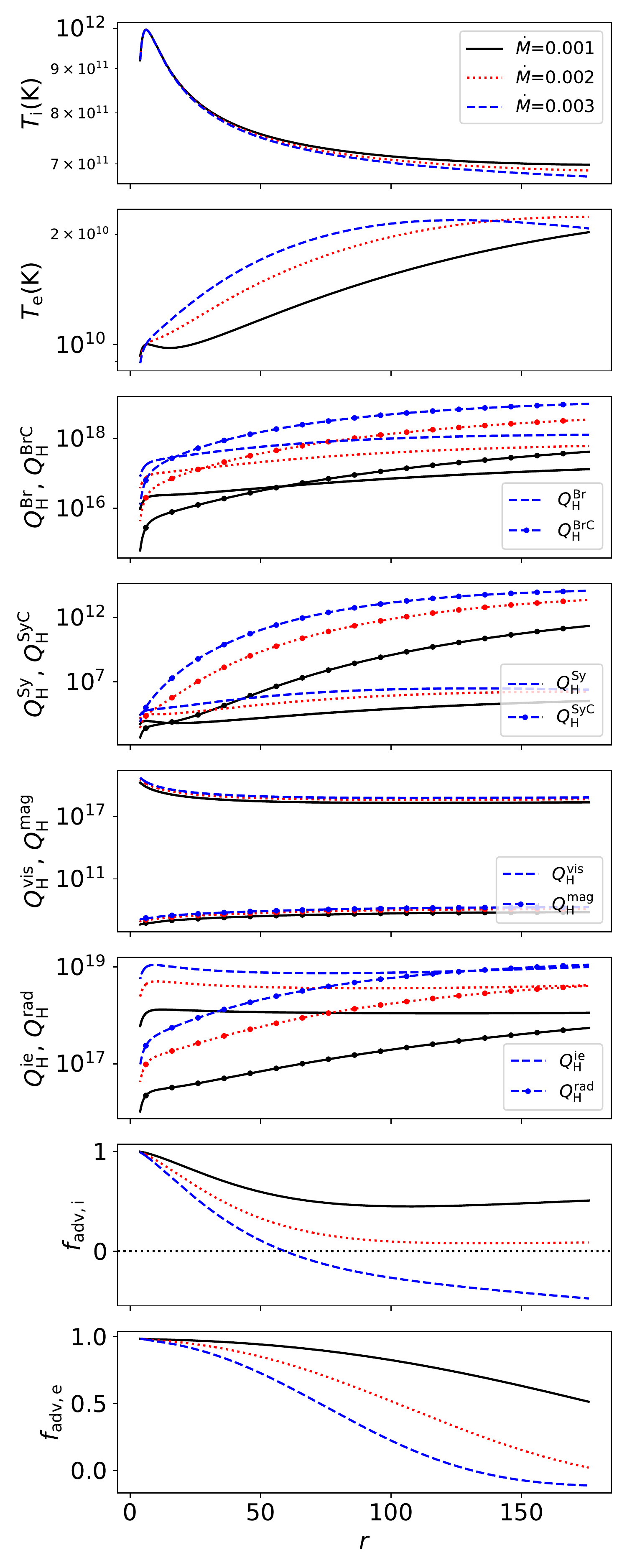}
\caption{Variation of physical variables related to thermal properties of the disc with accretion rate. All height integrated heating and cooling are in units of ergs cm$^{-2}$ sec$^{-1}$. This is for non-magnetic or very weakly magnetic case $(\text{f}_\mathrm{rc} = \text{f}_\mathrm{\phi c} = \text{f}_\mathrm{zc}=10^6)$ with $\alpha$=0.02. For $\dot{M}$=0.003 $\dot{M}_\mathrm{Edd}$, in the outer region, the disc becomes thermally unstable as the advection factor for ion (f$_{\text{adv,i}}$) becomes negative.}
\label{fig_mdot_effect_thermal_variables}
\end{figure}

In this section, we revisit the old-established result for non-magnetic or very weakly magnetic case: advection decreases with increasing accretion rate, which is shown in Fig. \ref{fig_mdot_effect_thermal_variables}. Magnetic field is kept negligible by setting f$_\mathrm{rc}$=f$_\mathrm{\phi c}$=f$_\mathrm{zc}=10^6$, leading to negligible magnetic heating, Synchrotron cooling and its comptonization. With increasing accretion rate, naturally density increases, subsequently the viscous dissipation, Coulomb coupling, and radiation from the disc increases. As the Coulomb coupling increases more than the viscous dissipation for the same increment in density, ion temperature decreases with the increment in accretion rate. However, the evolution of electron temperature depends very sensitively on the choice of electron temperature at the critical point.

In this scenario, primary cooling is governed by bremsstrahlung and its comptonization. We find that with increasing accretion rate, comptonization of bremsstrahlung increases largely and for $\dot{M}=0.003 \dot{M}_\mathrm{Edd}$, comptonization of bremsstrahlung dominates over bremsstrahlung itself for most of the radial range. As the density increases, the scattering probability of the soft photon increases largely, and for a higher accretion rate, most of the photons produced due to bremsstrahlung radiation are comptonized.

We present advection factor for ion and electron as $f_\mathrm{adv,i}$ and $f_\mathrm{adv,e}$ respectively, given by
\begin{equation}
f_\mathrm{adv,i}=\frac{Q^\mathrm{vis}_\mathrm{H}+Q^\mathrm{mag}_\mathrm{H}-Q^\mathrm{ie}_\mathrm{H}}{Q^\mathrm{vis}_\mathrm{H}+Q^\mathrm{mag}_\mathrm{H}},
\end{equation} 
and
\begin{equation}
f_\mathrm{adv,e}=\frac{Q^\mathrm{ie}_\mathrm{H}-Q^\mathrm{rad}_\mathrm{H}}{Q^\mathrm{ie}_\mathrm{H}}.
\end{equation}

We find that in the outer region, advection is less, which increases gradually in the inner region, and $f_\mathrm{adv,i}$ and $f_\mathrm{adv,e}$ tend to the value of unity in the innermost region. This result is also similar as found in earlier works (\citealt{Nakamura1997, Oda2012, Yuan2014}). To investigate the thermal stability, we focus on the advection factor of ions as the advected heat energy by ions plays a major role in thermally stabilizing the advective flows. We find that for our chosen parameter space with $\dot{M}=0.003\dot{M}_\mathrm{Edd}$, in the outer region of the disc $f_\mathrm{adv,i}$ becomes negative as shown in Fig. \ref{fig_mdot_effect_thermal_variables}. This is because with increasing $\dot{M}$, hence $\rho$, cooling of ions ($Q^\mathrm{ie}_\mathrm{H}$) increases faster than its total heating ($Q^\mathrm{vis}_\mathrm{H}+Q^\mathrm{mag}_\mathrm{H}$). Thus $Q^\mathrm{ie}_\mathrm{H}$ dominates over $Q^\mathrm{vis}_\mathrm{H}+Q^\mathrm{mag}_\mathrm{H}$, and advection can not act as a cooling for such higher accretion rate. It indicates that for $\dot{M}=0.003\dot{M}_\mathrm{Edd}$ the disc is thermally unstable. We can denote this as a critical accretion rate, $\dot{M}_\text{cr}$. With our chosen parameter space, $\dot{M}_\text{cr}$ indicates where the instability just kicks in. For the critical accretion rate or above, in the absence of magnetic field, the optically thin disc becomes thermally unstable.

\subsection{Thermal stabilization through strong large-scale magnetic field}
\label{section LSMF stabilizing disc}
\begin{figure}
\includegraphics[width=\columnwidth]{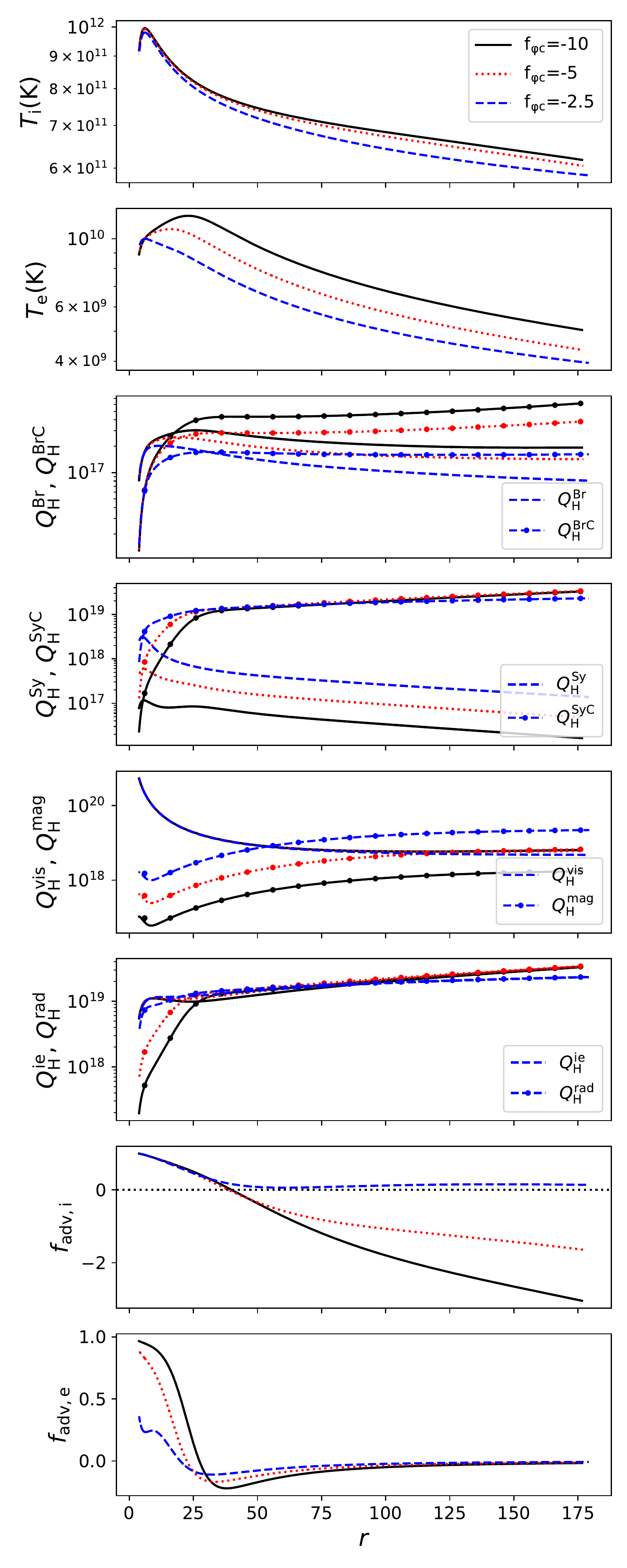}
\caption{Variation of various variables related to thermal stability of the disc with different magnetic field strength when $\dot{M}$=0.003 $\dot{M}_\mathrm{Edd}$ is fixed. Magnetic field configuration as well as all other parameters except $\dot{M}$ are same as in Fig.(\ref{fig_mag_effect_vr_cs}). All height integrated heating and cooling are presented in units of ergs cm$^{-2}$ sec$^{-1}$.}
\label{strong field stabilize the disc}
\end{figure}

The effect of the LSMF, as well as accretion rate on different physical variables related to the thermal stability of the disc, is shown in section \ref{section mag effect thermal} and section \ref{section instability due to mdot} respectively. In section \ref{section instability due to mdot} we see that with increasing accretion rate, thermal instability kicks in and in the absence of magnetic field, at or above the critical accretion rate ($\dot{M}_\text{cr}=0.003 \dot{M}_\mathrm{Edd}$), the optically thin disc becomes thermally unstable. In this section, we show that if strong LSMF is present, disc with $\dot{M}=\dot{M}_\text{cr}$ can regain its thermal stability. Here also as described in section \ref{section_result_suitable_magnetic_field_configuration}, the configuration of the magnetic field is fixed, which supports the $\alpha$-viscosity in outward transport of angular momentum.

The effect of the LSMF on different variables of the disc with $\dot{M}=\dot{M}_\text{cr}$ is shown in Fig. \ref{strong field stabilize the disc}. As discussed in section \ref{section instability due to mdot}, for the thermal stability of the disc, we focus on the positivity of the $f_\mathrm{adv,i}$. With the increasing strength of magnetic field, $Q^\mathrm{mag}_\mathrm{H}$ increases and dominates over $Q^\mathrm{vis}_\mathrm{H}$ in the outer region of the disc for f$_\mathrm{\phi c}>-5$ (or magnitude of f$_\mathrm{\phi c}<5$). We find that for f$_\mathrm{\phi c}=-2.5$, the magnetic heating serves as the main contributor in total heating in the outer region of the disc and plays a crucial role in making the $f_\mathrm{adv,i}$ positive. We find that $Q^\mathrm{ie}_\mathrm{H}$, which cools the ions and effectively decreases the $f_\mathrm{adv,i}$, remain almost the same with the increasing strength of the magnetic field. However, once $Q^\mathrm{mag}_\mathrm{H}$ is significant for f$_\mathrm{\phi c}=-2.5$, the significant increment in $Q^\mathrm{mag}_\mathrm{H}$ makes the $f_\mathrm{adv,i}$ positive throughout the whole radial range. Writing in physical units, a magnetic field with strength $5\times10^5-10^6$ Gauss is required to stabilize the disc for the parameter space we explore.

The evolution of $f_\mathrm{adv,e}$ is quite complex, sensitively depend on the choice of electron temperature at critical point ($T_\mathrm{ec}$) and remains negative for some radial range for our investigated strength of magnetic field. For this reason, we have done the analysis for three different $T_\mathrm{ec}$ values i.e. $5\times10^9$ K, $10^{10}$ K (for which plots are shown), and $5\times10^{10}$ K. We find that although the evolution of $T_\mathrm{e}$, $Q^\mathrm{ie}_\mathrm{H}$, $Q^\mathrm{rad}_\mathrm{H}$ and $f_\mathrm{adv,e}$ change significantly, $f_\mathrm{adv,i}$ remains positive for all the cases when f$_\mathrm{\phi c}=-2.5$ with $\dot{M}=0.003\dot{M}_\mathrm{Edd}$. It emphasizes that the main ingredient to make $f_\mathrm{adv,i}$ positive is $Q^\mathrm{mag}_\mathrm{H}$, not any contribution from $Q^\mathrm{ie}_\mathrm{H}$.

\begin{figure}
\includegraphics[width=\columnwidth]{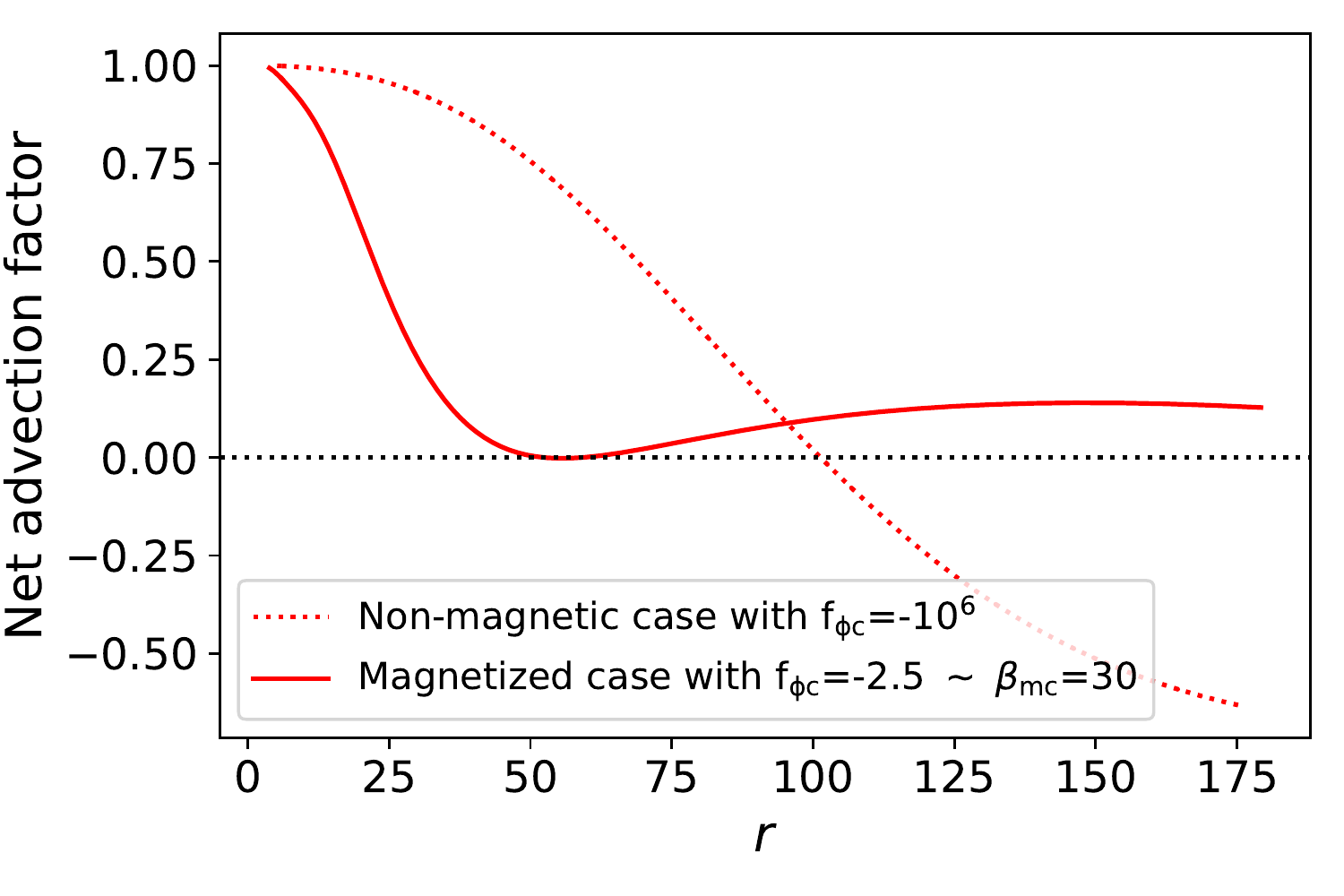}
\caption{Net advection factor ((total heating-radiative cooling)/total heating) for magnetized and non-magnetized case when $\dot{M}$=$\dot{M}_\text{cr}$ (0.003 $\dot{M}_\mathrm{Edd}$). $\alpha$=0.02 is fixed for all the calculations.}
\label{fig net advection factor}
\end{figure}

Although $f_\mathrm{adv,i}$ becomes positive with f$_\mathrm{\phi c}=-2.5$ and $\dot{M}=\dot{M}_\text{cr}$, $f_\mathrm{adv,e}$ remains negative for some radial range. To be confirm more strongly about the thermal stability of the flow, we also check the net advection factor ($f_\mathrm{adv}$). $Q^\mathrm{ie}_\mathrm{H}$ is the intermediate factor through which ions transfer energy to electrons. The accretion flow heats up through contributions from $Q^\mathrm{vis}_\mathrm{H}$, $Q^\mathrm{mag}_\mathrm{H}$ and cools down through $Q^\mathrm{rad}_\mathrm{H}$. We define the net advection factor as  
\begin{equation}
f_\mathrm{adv}=\frac{Q^\mathrm{vis}_\mathrm{H}+Q^\mathrm{mag}_\mathrm{H}-Q^\mathrm{rad}_\mathrm{H}}{Q^\mathrm{vis}_\mathrm{H}+Q^\mathrm{mag}_\mathrm{H}}.
\end{equation}
Fig. \ref{fig net advection factor} shows the radial variation of net advection factor for magnetic (f$_\mathrm{\phi c}=-2.5$) and non-magnetic or very weakly magnetic (f$_\mathrm{\phi c}=-10^6$) case while $\dot{M}=\dot{M}_\text{cr}$. It also confirms that whatever be the intermediate heat-transfer ($Q^\mathrm{ie}_\mathrm{H}$) becomes, a positive net heat is advected inward and makes the system thermally stable. We see that for f$_\mathrm{\phi c}=-2.5$, the net advection factor becomes just positive for the whole radial range, touches the zero line at some intermediate radius value. That is why it will be safe to state that for $\dot{M}=\dot{M}_\text{cr}$, disc becomes stable for f$_\mathrm{\phi c}>-2.5$ (magnitude of f$_\mathrm{\phi c}<2.5$). For $\dot{M}=\dot{M}_\text{cr}$, magnetic field corresponding to f$_\mathrm{\phi c}=-2.5$ acts as a critical value of the magnetic field above which disc is thermally stable.

\subsection{Outward transport of angular momentum through strong vertical magnetic field}
\label{section_vertical_field_transport}

\begin{figure}
\includegraphics[width=\columnwidth]{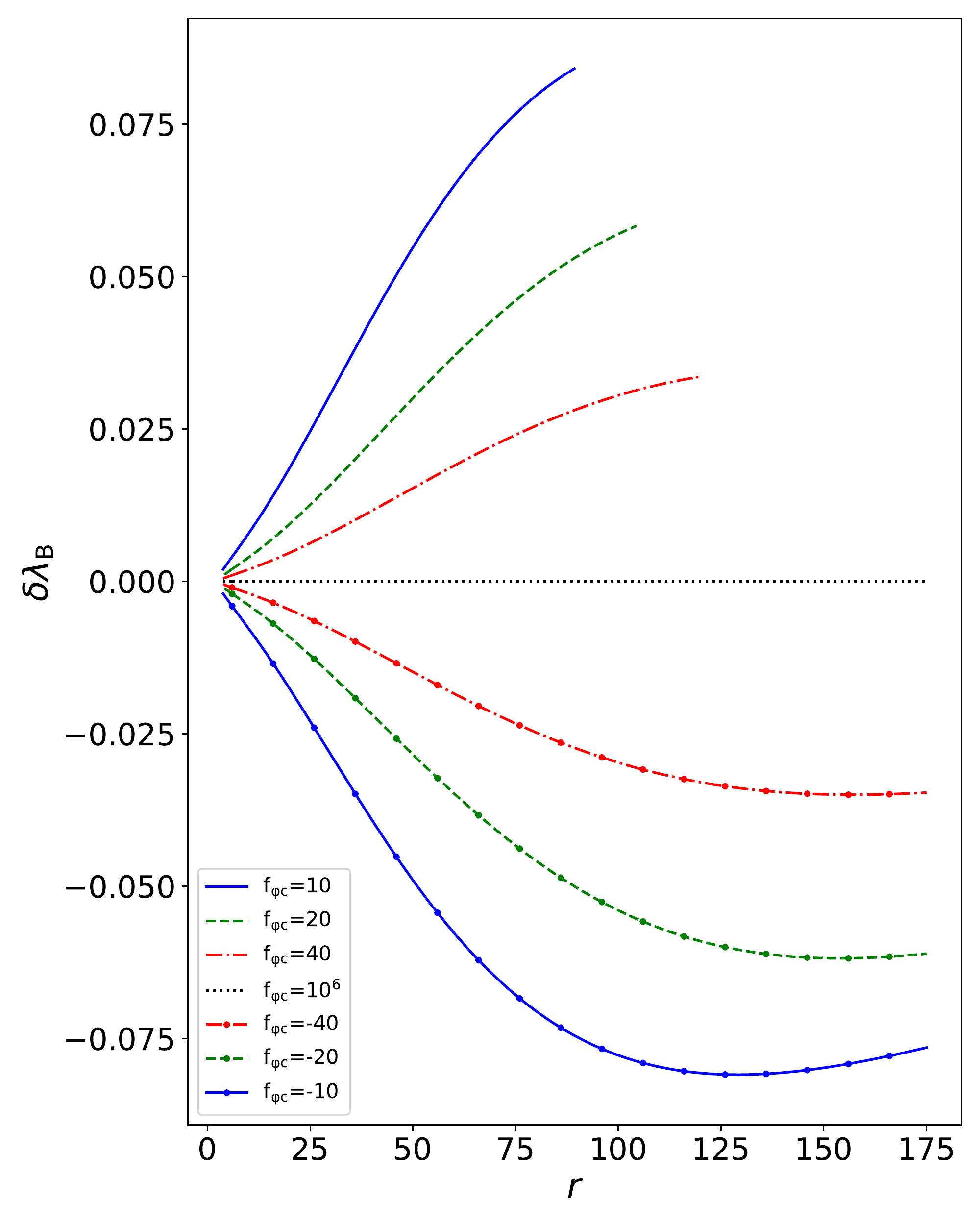}
\vspace{-0.5cm}
\caption{Magnetic contribution in transport of specific angular momentum ($\lambda$) for different magnitudes and orientation of $B_\mathrm{\phi}$ with strong vertical field. Positive value of $\delta\lambda_\mathrm{B}$ denotes the positive outward transport of angular momentum through the LSMF. $\alpha$ parameter is set to 0.02. The magnitude of magnetic field is set by f$_\mathrm{\phi c}$ value. Larger the f$_\mathrm{\phi c}$ value is, weaker the magnetic field is. $B_\mathrm{r}$ and $B_\mathrm{z}$ are in $\hat{r}$ and $\hat{z}$ direction always. $B_\mathrm{r}$ is kept very weak by fixing f$_\mathrm{rc}$=10$^6$ and strong vertical field is set by f$_\mathrm{zc}$=1.0. We fix $\dot{M}$=0.001 $\dot{M}_\mathrm{Edd}$.}
\label{fig_transport_through_vertical_field}
\end{figure}

We already discussed in section \ref{section_result_suitable_magnetic_field_configuration} that how positive value of $\delta\lambda_\mathrm{B}$ (equation \ref{equation_dlambda/dr_B}) is possible when $B_\mathrm{r}B_\mathrm{\phi}$ is negative and $B_\mathrm{z}B_\mathrm{\phi}$ is positive. Till now, we have found only the effectiveness of strong $B_\mathrm{r}B_\mathrm{\phi}$ in the disc and thermal stabilization of the disc with the help of that field. In this section, we present that with negligible $B_\mathrm{r}$, a strong vertical field in positive $\hat{z}$ can also efficiently transport angular momentum outward through dominant $B_\mathrm{z}B_\mathrm{\phi}$, and help turbulent viscosity in the formation of the disc. As we discussed earlier, we are solving in the upper half-plane of the disc, $B_\mathrm{r}$ and $B_\mathrm{z}$ are fixed in positive $\hat{r}$ and positive $\hat{z}$ direction. So, to make $B_\mathrm{z}B_\mathrm{\phi}$ positive, $B_\mathrm{\phi}$ has to be in positive $\hat{\phi}$ direction. Again with this configuration, $B_\mathrm{r}B_\mathrm{\phi}$ also will become positive and oppose the outward transport. That is why with a strong vertical field, the radial field should be very weak to make the significant outward transport possible. We keep $B_\mathrm{r}$ very low by setting f$_{rc}=10^6$. In Fig. \ref{fig_transport_through_vertical_field}, we show the contribution of the LSMF in outward transport of angular momentum, i.e., $\delta\lambda_\mathrm{B}$ when there is a strong vertical field present in the disc with the negligible radial field. Positive and negative values of f$_\mathrm{\phi c}$ represent that $B_\mathrm{\phi}$ is in positive and negative $\hat{\phi}$ direction respectively. We make the vertical field strong by keeping f$_{zc}=1.0$ and varying the strength of $B_\mathrm{\phi}$. When $B_\mathrm{\phi}$ is positive, a stronger field increases the value of $\delta\lambda_\mathrm{B}$, transports angular momentum more efficiently, and helps $\alpha$-viscosity in outward transport as well as in the formation of the disc. Again, when $B_\mathrm{\phi}$ has reversed to negative $\hat{\phi}$ direction, a stronger field does inward transport of angular momentum and is shown in the lower half of Fig. \ref{fig_transport_through_vertical_field} where $\delta\lambda_\mathrm{B}$ is negative. The Black dotted line with $\delta\lambda_\mathrm{B}=0$ represents the non-magnetic or very weakly magnetic case where only turbulent viscosity transports angular momentum outward and forms the disc. This magnitude of the magnetic field is comparable to the field value when transport occurs through $B_\mathrm{r}B_\mathrm{\phi}$, and it will produce Joule heating of the same value. Hence, we can state that a strong toroidal field with a weak vertical field or a strong vertical field with a weak radial field can transport the angular momentum outward and thermally stabilize the disc.

It is interesting to see that when the positive transport occurs, the size of the disc decreases with the increasing strength of the LSMF. This is because with the stronger magnetic field, the magnetic contribution in outward transport of angular momentum increases and dominates over turbulent transport. Therefore, the faster outward transport of angular momentum with the increasing strength of magnetic field leads to smaller size of the disc. However, the dependency of the outer boundary of the disc on the LSMF comes into effect only when the strength of the magnetic field is significant and the outward transport of angular momentum through the LSMF dominates over $\alpha$-viscosity. That is why for toroidally dominated LSMF case (Fig. \ref{fig mag helping opposing transport}), the size of the disc does not depend on the strength of the magnetic field as the magnetic field is relatively weaker and turbulent transport remains the dominating mechanism for transport. However, it is worth to mention that after comparing critically, we find that vertical field dominated case is slightly more efficient in transporting angular momentum outward compared to toroidally dominated case for our investigated parameter regime.

\section{Discussion}
\label{section Discussion}
\begin{figure}
\includegraphics[width=\columnwidth]{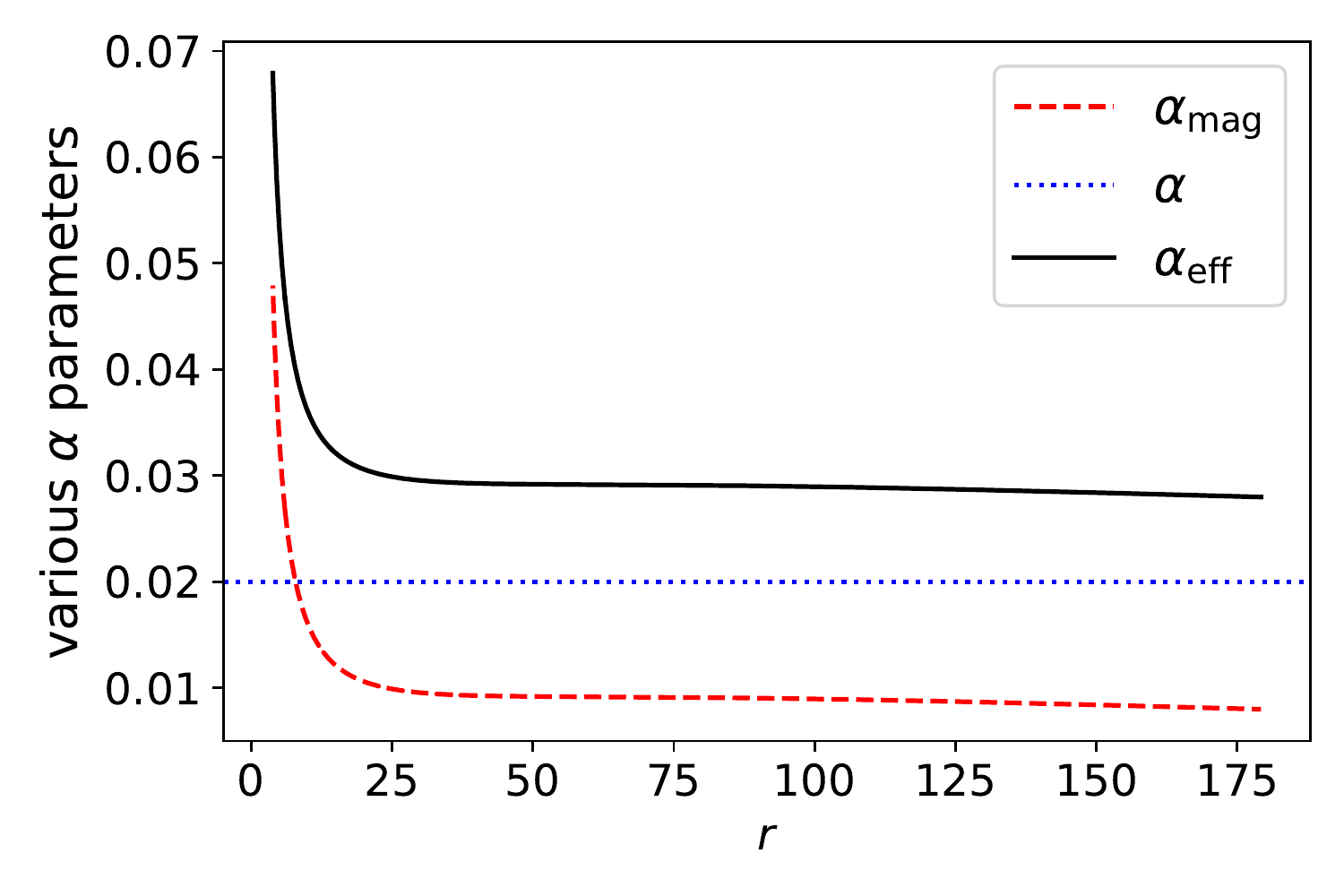}
\caption{Various $\alpha$ parameters when the LSMF just thermally stabilize the disc with f$_\mathrm{\phi c}$=-2.5 and $\dot{M}$=$\dot{M}_\text{cr}$ (0.003 $\dot{M}_\mathrm{Edd}$).}
\label{fig alpha parameters}
\end{figure}

We see that the LSMF can support or oppose the $\alpha$-viscosity in outward transport of angular momentum depending on its configuration. If we ignore the negligible contribution from $B_\mathrm{z}$ for our usual (toroidally dominated magnetic field) case, from equation \eqref{azimuthal component NS} and also from equation (14) of \cite{Jacquemin-Ide2021}, we can infer that the shearing stress related to the LSMF is $B_\mathrm{r}B_\mathrm{\phi}/4\pi$ whereas $\sigma_\mathrm{r\phi}^{'}$ is the shearing stress of turbulent origin. Now, whether the LSMF will support the turbulent transport or not that depends on the positive value of $\delta\lambda_\mathrm{B}$, which again depends on the direction and strength of different components of the magnetic field as described in section \ref{section_result_suitable_magnetic_field_configuration}. Radial velocity $v_\mathrm{r}$ in negative $\hat{r}$ direction indicates successful accretion. That is why all the values of $v_\mathrm{r}$ presented in this work have negative values. Now $\dot{M}$ is always positive and correspondingly $\dot{M}=-4\pi N_1r\rho Hv_\mathrm{r}$. The numerical constant $N_1$ appears due to the vertical integration of $\rho$. The negative sign present in the expression for $\dot{M}$ leads to the turbulent stress $\sigma_\mathrm{r\phi}^{'}=-\alpha(p+\rho v_\mathrm{r}^2)$. This sign convention of turbulent transport must be taken care of appropriately as it will dictate the configuration of the magnetic field, which will help $\alpha$-viscosity in transporting angular momentum outward. $\alpha$ is the proportionality constant of shear stress of turbulent origin. The LSMF with suitable configuration contributes to transport approximately by -$B_\mathrm{r}B_\mathrm{\phi}$/4$\pi$ along with turbulent stress, which finally leads to net transport. Hence, we find effective $\alpha$-viscosity $\alpha_\mathrm{eff}$ as follows:
\begin{equation*}
	\alpha_\mathrm{mag}=\frac{1}{p}\Big(-\frac{B_\mathrm{r}B_\mathrm{\phi}}{4\pi}\Big), \text{ }  \alpha_\mathrm{eff}=\alpha+\alpha_\mathrm{mag}.
\end{equation*}
For the case which stabilizes the disc with accretion rate of $\dot{M}_\text{cr}$ and f$_\mathrm{\phi c}=-2.5$, different $\alpha$-parameters are shown in Fig. \ref{fig alpha parameters}. We can see that the effective value of $\alpha$ lies within 0.03-0.07. This range is at the lower side of the effective value of $\alpha$, predicted by the MRI simulation threaded by the LSMF (\citealt{Lesur2013, Bai2013, Salvesen2016}). These simulations showed that $\alpha$ increases with decreasing $\beta_\mathrm{m}$ and could reach up to unity. However, the observed range of $\alpha$-value during outburst for 12 black hole low mass X-ray binaries is 0.2-1.0, and strong disc winds can also influence the inferred $\alpha$-value from observation (\citealt{Tetarenko2018}). Keep in mind that the magnetic field, which we are using to find $\alpha_\mathrm{eff}$, is at the minimum value which is required to stabilize the disc thermally. Correspondingly, the presented $\alpha_\mathrm{eff}$ is the minimum possible value in the disc at $\dot{M}=\dot{M}_\text{cr}$. A similar treatment regarding the role of large-scale magnetic stress over the turbulent viscous stress has been discussed by \cite{Mondal2019, Mondal2020} in the context of disc-outflow symbiosis with much stronger LSMF.

It is also quite interesting that by reversing $B_\mathrm{\phi}$, the LSMF can oppose the outward transport of angular momentum. Even a strong enough magnetic field with positive $B_\mathrm{r}B_\mathrm{\phi}$ or negative $B_\mathrm{z}B_\mathrm{\phi}$ can resist the outward transport through turbulent $\alpha$-viscosity and halts the formation of the disc. Inward transport of angular momentum through the LSMF, i.e., the negative value of $\delta\lambda_{B}$, is shown in the bottom panel of Figs. \ref{fig mag helping opposing transport} and \ref{fig_transport_through_vertical_field}.

Another exciting outcome is that the results of the whole analysis remain the same if $\boldsymbol{B}$ changes to -$\boldsymbol{B}$, then only the components of the magnetic field change their sign. This indicates that the disc dynamics and the heating and cooling of the disc are symmetric on the reversal of the magnetic field. Taking all the possibilities investigated in this work, in Table \ref{table_configuration_of_magnetic_field} we present different configurations of the magnetic field which support or oppose the $\alpha$-viscosity in outward transport of angular momentum.

\begin{table*}
	\centering
	\caption{Configuration of magnetic field which supports and opposes outward transport of angular momentum.}
	\label{table_configuration_of_magnetic_field}
	\begin{tabular}{|c|c|c|c|}
	\hline
	$B_\mathrm{r}$ & $B_\mathrm{\phi}$ & $B_\mathrm{z}$ & Effect in outward transport\\
	\hline
	Large magnitude in positive $\hat{r}$ & Larger magnitude in positive $\hat{\phi}$ & Small magnitude in positive $\hat{z}$ & Oppose $\alpha$-viscosity\\
	\hline
	Large magnitude in positive $\hat{r}$ & Larger magnitude in negative $\hat{\phi}$ & Small magnitude in positive $\hat{z}$ & Support $\alpha$-viscosity\\
	\hline
	Large magnitude in negative $\hat{r}$ & Larger magnitude in negative $\hat{\phi}$ & Small magnitude in negative $\hat{z}$ & Oppose $\alpha$-viscosity\\
	\hline
	Large magnitude in negative $\hat{r}$ & Larger magnitude in positive $\hat{\phi}$ & Small magnitude in negative $\hat{z}$ & Support $\alpha$-viscosity\\
	\hline
	Small magnitude in positive $\hat{r}$ & Large magnitude in positive $\hat{\phi}$ & Large or larger magnitude in positive $\hat{z}$ & Support $\alpha$-viscosity\\
	\hline
	Small magnitude in positive $\hat{r}$ & Large magnitude in negative $\hat{\phi}$ & Large or larger magnitude in positive $\hat{z}$ & Oppose $\alpha$-viscosity\\
	\hline
    Small magnitude in negative $\hat{r}$ & Large magnitude in negative $\hat{\phi}$ & Large or larger magnitude in negative $\hat{z}$ & Support $\alpha$-viscosity\\
	\hline
	Small magnitude in negative $\hat{r}$ & Large magnitude in positive $\hat{\phi}$ & Large or larger magnitude in negative $\hat{z}$ & Oppose $\alpha$-viscosity\\
	\hline
	\end{tabular}
\end{table*}

Calculation from micro-physics (various channels for electron heating such as current-driven plasma instabilities, turbulence in the presence of the magnetic field, and pressure anisotropy are discussed in \citealt{BisnovatyiKogan1997, Quataert1999, Sharma2007, Lehe2009}) estimates $\delta\gg10^{-2}$. Observational modeling of low-luminous active galactic nuclei (\citealt{Yu2011, Liu2013}) gives its value $\approx$ 0.1. Hence, there is no consensus about the exact value of $\delta$. For simplicity, we fix $\delta=0$ throughout this work. Also, the disc is heated through two channels: viscous dissipation and Joule heating, and possibly different fractions of these two heatings directly go to the electron. As we mentioned earlier, $\delta$ parameter is crucial for the evolution of $T_\mathrm{e}$, i.e., for modeling the spectra. For $\delta \approx$ 0.1, ions will be heated less, $\dot{M}_\text{cr}$ will be lower than 0.003 $\dot{M}_\mathrm{Edd}$, i.e. instability will kick in at a lower $\dot{M}$. As $\dot{M}_\text{cr}$ will become small, the weaker magnetic field will be able to stabilize the disc thermally. This indicates that the inclusion of the non-zero $\delta$ will only change $\dot{M}_\text{cr}$ and the value of the required magnetic field for stabilization; the main conclusion will remain the same. In addition to $\delta$, $Q^\mathrm{ie}_\mathrm{H}$ and $Q^\mathrm{rad}_\mathrm{H}$ sensitively depend on the choice of $T_\mathrm{ec}$. However, redoing the calculations with different $T_\mathrm{ec}$ values although change the evolution of $Q^\mathrm{ie}_\mathrm{H}$ and $Q^\mathrm{rad}_\mathrm{H}$ significantly, the minimum magnetic field strength required to stabilize the disc remains the same.

The above discussions converge to the point that $Q^\mathrm{ie}_\mathrm{H}$ or $Q^\mathrm{rad}_\mathrm{H}$ has no contribution in stabilization. The rising of Joule heating with the increasing strength of the magnetic field plays a key role in making the advection factor positive and stabilizing the disc thermally. The unstable optically thin flow at or above $\dot{M}_\text{cr}$ regains its stability due to the addition of Joule heating in the system through the LSMF. This is quite similar to the fact about the dependence of $\dot{M}_\text{cr}$ on the Shakura-Sunyaev $\alpha$-parameter in a non-magnetic or weakly magnetic case ($\dot{m}_\mathrm{max}$ in \citealt{Abramowicz1995}). As $\alpha$-viscosity increases, heating increases and making the unstable advective accretion disc stable for a fixed accretion rate. In the present work, $\alpha_{eff}$ plays the same role as $\alpha$-viscosity mentioned above.

\section{Caveats}
\label{section caveats}
In our model, $\lambda=\lambda_\mathrm{K}$ is considered to be the outer boundary of the disc to mimic the situation that in the outer region of the hot advective disc, a cold Keplerian disc exists. This truncated disc geometry is useful in explaining simultaneous observation of soft and hard X-rays from the same source. We know that gravity almost balances the centrifugal force, and force due to pressure gradient is practically zero for a Keplerian disc. However, after solving the equation for sub-Keplerian flow with $\lambda=\lambda_\mathrm{K}$ as the outer boundary condition, we have found that at the outer boundary, gravity does not balance the centrifugal force and force due to pressure gradient remains significant. These things indicate that the transition from sub-Keplerian to Keplerian flows is abrupt though $\lambda=\lambda_\mathrm{K}$ and this boundary layer problem remains an open area to address.

We have assumed the disc to be vertically isothermal, which may not be the case in reality. Also, to handle the second derivative and square terms present in the induction equation and Joule heating, we have approximated as the following: $d^2B_\mathrm{\phi}/dr^2 = 2B_\mathrm{\phi}/r^2$, $d^2B_\mathrm{z}/dr^2 = 2B_\mathrm{z}/r^2$, and $(dB_\mathrm{\phi}/dr)^2 = B_\mathrm{\phi}^2/r^2$, $(dB_\mathrm{z}/dr)^2 = B_\mathrm{z}^2/r^2$. Although replacing these derivatives by other values will change the results, we believe that the main conclusion will remain the same qualitatively. Again, we have restricted our parameter space such that $B_\mathrm{\phi}$ always remains unidirectional throughout the whole radial range. Nevertheless, allowing the reversing of $B_\mathrm{\phi}$ with radial distance will broaden the parameter space as well as can lead to many diverse results.

\section{Summary}
\label{section summary}
In this work, we have explored the possible contribution of the LSMF in transporting angular momentum outward in addition to turbulent transport, which is here based on the Shakura-Sunyaev $\alpha$-parameter. We have also explicitly explored the magnetic field's effect in heating and cooling the disc. The key findings are summarized below.
\begin{itemize}
	\item{Depending on the magnetic field configuration, it supports/opposes the turbulent contribution in outward transport. For the toroidally dominated field, the vertical field should be very weak to make transport efficient. Different configurations and their effects are presented in Table \ref{table_configuration_of_magnetic_field}.}
\end{itemize}
\begin{itemize}
	\item{Configuration of the LSMF is fixed to meet the requirement of outward transport of angular momentum. With the increasing strength of the toroidal field, although the magnetic contribution in outward transport increases, the disc becomes cooler, and turbulent transport decreases.}
\end{itemize}
\begin{itemize}
	\item{In the absence of magnetic field, naturally, with increasing accretion rate, advection of heat decreases. For $\dot{M}$ $\gtrsim$ $\dot{M}_\text{cr}$ ($\dot{M}_\text{cr}= 0.003 \dot{M}_\mathrm{Edd}$ for our chosen parameter space) advected heat energy through ions becomes negative and disc becomes thermally unstable.}
\end{itemize}
\begin{itemize}
	\item{Disc with $\dot{M}\gtrsim\dot{M}_\text{cr}$ regains its thermal stability in the presence of a strong enough magnetic field. Joule heating plays a crucial role in stabilizing the disc. The LSMF with a suitable configuration transports angular momentum outward as well as stabilizes the optically thin disc with $\dot{M}\gtrsim\dot{M}_\text{cr}$.}
\end{itemize}
\begin{itemize}
	\item{A strong vertical field with a weak radial field also has an equal potential to help $\alpha$-viscosity in outward transportation of angular momentum and stabilize the disc thermally.}
\end{itemize}
\begin{itemize}
	\item{For the most magnetically dominated case in our analysis, the value of $\beta_\mathrm{m}$ lies within the range of 5-10$^3$. This confirms that there is no restriction for simultaneous operation of the LSMF and the $\alpha$-viscosity even if we assume the MRI solely to be the source of turbulent $\alpha$-viscosity.}
\end{itemize}

\section*{Acknowledgements}

SRD would like to thank Jonathan Ferreira, Susmita Chakravorty, Prasun Dhang, and Subham Ghosh for their formal and informal scientific discussions. The authors are also thankful to the anonymous referee for insightful questions which made the manuscript much better. This work is partly supported by the fund of DST INSPIRE fellowship belonging to SRD and partly by the project with research Grant No. DSTO/PPH/BMP/1946 (EMR/2017/001226) belonging to BM.

\section*{Data Availability}
Authors will be happy to share the data presented in this work on a reasonable request.


\bibliographystyle{mnras}
\bibliography{mhd_adv} 




\appendix

\section{Final height averaged equations}
\label{Appendix_final_equations}
After averaging equations (\ref{continuity}) - (\ref{no monopole equation}) vertically from 0 to $H$, we obtain finally nine coupled ordinary differential equations as follows:
\begin{equation}\label{averaged continuity}
H\frac{d}{dr}(r\rho_0v_\mathrm{r})+\frac{1}{3}(r\rho_0v_\mathrm{r})\Big(\frac{dH}{dr}\Big)=0,
\end{equation}
\begin{multline}\label{averaged radial momentum}
v_\mathrm{r}\frac{dv_\mathrm{r}}{dr}-\frac{v^2_\mathrm{\phi}}{r}+\frac{1}{\rho_0}\frac{\partial p_0}{\partial r}+\frac{p_0}{3\rho_0H}\frac{dH}{dr}+\frac{1}{4\pi\rho_0}\Bigg[\frac{B^2_\mathrm{\phi 0}}{r}\\
+B_\mathrm{\phi 0}\left(\frac{dB_\mathrm{\phi 0}}{dr}+\frac{B_\mathrm{\phi 0}}{6H}\frac{dH}{dr}\right)+N_3B_\mathrm{z}\frac{dB_\mathrm{z}}{dr}\Bigg]-F_\mathrm{r}(r)=0,
\end{multline}
\begin{multline}\label{final equation angular momentum transport}
v_\mathrm{r}\frac{dv_\mathrm{\phi}}{dr}+\frac{v_\mathrm{r}v_\mathrm{\phi}}{r}+\frac{\alpha}{\rho_0}\frac{dp_0}{dr}+2\alpha v_\mathrm{r}\frac{dv_\mathrm{r}}{dr}+\alpha\frac{v^2_\mathrm{r}}{\rho_0}\frac{d\rho_0}{dr}+\\
\frac{2\alpha p_0}{r\rho_0}+\frac{2\alpha v^2_\mathrm{r}}{r}+\frac{\alpha}{\rho_0}(p_0+\rho_0v^2_\mathrm{r})\Big(\frac{1}{3H}\frac{dH}{dr}\Big)\\
=\frac{1}{4\pi\rho_0}\Bigg[N_4\left(B_\mathrm{r0}\frac{dB_\mathrm{\phi 0}}{dr}+\frac{B_\mathrm{r}B_\mathrm{\phi 0}}{r}\right)\\
+\frac{N_5B_\mathrm{r}B_\mathrm{\phi 0}}{2H}\frac{dH}{dr}+\frac{N_6B_\mathrm{z}B_\mathrm{\phi 0}}{H}\Bigg],
\end{multline}
\begin{multline}\label{averaged ion energy}
-N_1H\Big[\frac{3}{2}v_\mathrm{r}\frac{dp_0}{dr}+Av_\mathrm{r}\frac{p_0}{\rho_0}\frac{d\rho_0}{dr}+Bv_\mathrm{r}\frac{p_0}{T_\mathrm{e}}\frac{dT_\mathrm{e}}{dr}\Big]\\
-\frac{N_2}{4}\Big(3+2A\Big)v_\mathrm{r}p_0\Big(\frac{dH}{dr}\Big)=Q^\mathrm{vis}_\mathrm{H}+Q^\mathrm{mag}_\mathrm{H}-Q^\mathrm{ie}_\mathrm{H},
\end{multline}
\begin{multline}\label{averaged electron energy}
-N_1H\Big[C\frac{v_\mathrm{r}p_0}{T_\mathrm{e}}\frac{dT_\mathrm{e}}{dr}+D\frac{v_\mathrm{r}p_0}{\rho_0}\frac{d\rho_0}{dr}\Big]-(N_2/2)Dv_\mathrm{r}p_0\Big(\frac{dH}{dr}\Big)\\
=Q^\mathrm{ie}_\mathrm{H}-Q^\mathrm{rad}_\mathrm{H},
\end{multline}
\begin{equation}\label{averaged radial induction}
\eta_\mathrm{B}\left[\frac{1}{r}\frac{\partial}{\partial r}\left(r\frac{\partial B_\mathrm{r}}{\partial r}\right)-\frac{B_\mathrm{r}}{r^2}\right]=0,
\end{equation}
\begin{multline}\label{averaged azimuthal induction}
\Big(B_\mathrm{r}\frac{dv_\mathrm{\phi}}{dr}+v_\mathrm{\phi}\frac{dB_\mathrm{r}}{dr}\Big)H-\Big(B_\mathrm{\phi 0}\frac{dv_\mathrm{r}}{dr}+v_\mathrm{r}\frac{dB_\mathrm{\phi 0}}{dr}+\eta_\mathrm{B}\frac{B_\mathrm{\phi 0}}{r^2}\Big)N_9H-\\
(N_8/2)v_\mathrm{r}B_\mathrm{\phi 0}\frac{dH}{dr}+
\eta_\mathrm{B}\Big[\Big\{\frac{1}{r}\frac{dB_\mathrm{\phi 0}}{dr}+\alpha_2(1+\alpha_2)\frac{B_\mathrm{\phi 0}}{r^2}\Big\}N_9H\\
+N_{13}\frac{B_\mathrm{\phi 0}}{H}+N_8\frac{B_\mathrm{\phi 0}}{2r}\frac{dH}{dr}+N_8\Big(\frac{dB_\mathrm{\phi 0}}{dr}\Big)\Big(\frac{H}{r}\Big)\Big]=0,
\end{multline}
\begin{equation}\label{averaged vertical induction}
-B_\mathrm{z}\frac{dv_\mathrm{r}}{dr}-v_\mathrm{r}\frac{dB_\mathrm{z}}{dr}-\frac{v_\mathrm{r}B_\mathrm{z}}{r}+\frac{\eta_\mathrm{B}}{r}\left[\frac{dB_\mathrm{z}}{dr}+\alpha_1(1+\alpha_1)\frac{B_\mathrm{z}}{r}\right]=0.
\end{equation}
\begin{equation}\label{averaged no monopole}
\frac{dB_\mathrm{r}}{dr}=-\frac{B_\mathrm{r}}{r}.
\end{equation}
Since $\eta_\mathrm{B} \neq 0$, the equation (\ref{averaged radial induction}) is same as the equation (\ref{averaged no monopole}). Here, $Q^\mathrm{vis}_\mathrm{H}$ is the height integrated viscous heating, i.e., heat generated per unit area of the disc per unit time through turbulent shear, as given by
\begin{equation*}
Q^\mathrm{vis}_\mathrm{H}=\int_{0}^{H}Q^\mathrm{vis} dz=\int_{0}^{H}\frac{\boldsymbol{\sigma_\mathrm{\bf{ik}}^{'}}^2}{\eta_\mathrm{V}} dz=N_1H \alpha (p + \rho v_\mathrm{r}^2)(v_\mathrm{\phi}/r).
\end{equation*}
$Q^\mathrm{mag}_\mathrm{H}$ is the height integrated Joule heating, i.e., heat generated per unit area per unit time due to magnetic contribution, as given by
\begin{multline*}
Q^\mathrm{mag}_\mathrm{H}=\int_{0}^{H}Q^\mathrm{mag} dz=\int_{0}^{H}\frac{j^2}{\sigma} dz=\int_{0}^{H}\frac{\eta_\mathrm{B}}{4\pi}(\nabla\times\boldsymbol{B})^2 dz\\
=\frac{\eta_\mathrm{B}}{4\pi}\Big[N_2\frac{B^2_\mathrm{\phi 0}}{8H}+\alpha^2_1\frac{B^2_\mathrm{z}}{r^2}H+N_1H\Big\{(1+\alpha^2_2)\Big(\frac{B_\mathrm{\phi 0}}{r}\Big)^2+\frac{2B_\mathrm{\phi 0}}{r}\Big(\frac{dB_\mathrm{\phi 0}}{dr}\Big)\Big\}\\+
N_2\Big\{B_\mathrm{\phi 0}\Big(\frac{dB_\mathrm{\phi 0}}{dr}\Big)\Big(\frac{H}{2r}\Big)+\frac{B^2_\mathrm{\phi 0}}{2r}\Big(\frac{dH}{dr}\Big)\Big\}\Big].
\end{multline*}
Due to technical reasons, we have approximated $d^2B_\mathrm{\phi}/dr^2$, $d^2B_\mathrm{z}/dr^2$ present in the $\phi$- and $z$-components of induction equation, and $(dB_\mathrm{\phi}/dr)^2$, $(dB_\mathrm{z}/dr)^2$ present in the Joule heating expression. In order to do so, we assume 
\begin{multline*}
\frac{d^2B_\mathrm{z}}{dr^2}=\alpha_1(1+\alpha_1)\frac{B_\mathrm{z}}{r^2}, \ \frac{d^2B_\mathrm{\phi 0}}{dr^2}=\alpha_2(1+\alpha_2)\frac{B_\mathrm{\phi 0}}{r^2}, \text{and} \\ \left(\frac{dB_\mathrm{z}}{dr}\right)^2=\alpha_1^2\frac{B_\mathrm{z}^2}{r^2}, \left(\frac{dB_\mathrm{\phi}}{dr}\right)^2=\alpha_2^2\frac{B_\mathrm{\phi}^2}{r^2},\\
\end{multline*}
where $\alpha_1$, $\alpha_2$ are fixed to unity throughout our calculation.
The parameters $A$, $B$, $C$, and $D$ are defined as different combinations of $\beta$-parameters ($\beta_\mathrm{i}$, $\beta_\mathrm{e}$, and $\beta_\mathrm{m}$):
\begin{equation*}
A=-[(5/2)\beta_\mathrm{i}+(3/2)\beta_\mathrm{e}+(5/2)(\beta_\mathrm{i}/\beta_\mathrm{m})+(3/2)(\beta_\mathrm{e}/\beta_\mathrm{m})],
\end{equation*}
\begin{equation*}
B=-[6-6\beta_\mathrm{i}-6(\beta_\mathrm{i}/\beta_\mathrm{m})-(9/2)\beta_\mathrm{e}-(9/2)(\beta_\mathrm{e}/\beta_\mathrm{m})],
\end{equation*}
\begin{equation*}
C=(3/2)\beta_\mathrm{e}(1+1/\beta_\mathrm{m}),\ \text{and}\ D=-\beta_\mathrm{e}(1+1/\beta_\mathrm{m}),
\end{equation*}
where 
\begin{equation}
\label{definition beta}
\beta_\mathrm{i}=\frac{p_\mathrm{i}}{p+B^2/8\pi},\ \beta_\mathrm{e}=\frac{p_\mathrm{e}}{p+B^2/8\pi},\ \text{and}\ \beta_\mathrm{m}=\frac{p}{B^2/8\pi}.
\end{equation}
$\beta_\mathrm{i}$ and $\beta_\mathrm{e}$ denote the fraction of ion pressure and electron pressure to total pressure, respectively. Total pressure includes $p$ and the magnetic pressure of the system, whereas $p$ includes ion, electron and radiation pressures, as given in equation (\ref{eq:pressure}).
The values of all the numerical coefficients are given by
\begin{multline*}
N_1=\sqrt{\frac{\pi}{2}} \left[\mathrm{erf}\left(\frac{1}{\sqrt{2}}\right)\right]=0.855624,\\
N_2=-\frac{2}{\sqrt{e}}+\sqrt{2\pi} \left[\mathrm{erf}\left(\frac{1}{\sqrt{2}}\right)\right]=0.498187,\\
N_3=\sqrt{\frac{\pi}{2}} \left[\mathrm{erfi}\left(\frac{1}{\sqrt{2}}\right)\right]=1.19496,\\
N_4=\sqrt{\pi} \left[\mathrm{erfi}\left(\frac{1}{2}\right)\right]=1.08997,\\
N_5=2\left[e^{1/4}-\sqrt{\pi} \left\lbrace \mathrm{erfi}\left(\frac{1}{2}\right)\right\rbrace\right]=0.388102,\\
N_6=1-e^{1/4}=-0.284025,\\
N_8=-\frac{2}{e^{1/4}}+2\sqrt{\pi} \left[\mathrm{erf}\left(\frac{1}{2}\right)\right]=0.287522,\\
N_9=\sqrt{\pi} \left[\mathrm{erf}\left(\frac{1}{2}\right)\right]=0.922562,\\
N_{12}=\frac{\sqrt{\pi}}{2} \left[\mathrm{erf}(1)\right]=0.746824,\\
N_{13}=-\frac{1}{2e^{1/4}}=-0.3894.\\
\end{multline*}

\onecolumn
The expressions of numerator ($\mathcal{N}$) and denominator ($\mathcal{D}$) of $dv_\mathrm{r}/dr$ are as following:
\begin{multline} \label{eq:numerator}
\mathcal{N}=-(v_r ((150.80 p+6 B_{\phi }^2) (r v_r-\eta _B) (-(8 p \pi +B_{\phi }^2)\\(\frac{3 C H^2 r (-2+3 r) N_1 v_r}{-2+r}+\frac{2 (8 H \pi  r (B r (Q_H^{\text{ie}}-Q_H^{\text{rad}})+C (r Q_H^{\text{ie}}-H \alpha  N_1 (p+\rho  v_r^2) v_{\phi }))-2 C H^2 B_z^2 \eta _B-4 C B_{\phi }^2 (H^2 N_1+0.06 r^2 N_2) \eta _B)}{8 p \pi +B_{\phi }^2})\\ (r \alpha  B_r B_{\phi }+r B_r^2 N_4+\rho  v_r (-12.57 r N_9 v_r+12.57 N_8 \eta _B+12.57 N_9 \eta _B))\\-2 H^2 (12 C B_{\phi } (r \alpha  B_r (29.32 p+B_{\phi }^2+4.19 \rho  v_r^2)+B_{\phi } (-0.5 r B_r^2 N_5+\rho  N_8 v_r(6.28 r v_r-6.28 \eta _B)))(-0.75 r N_1 v_r+(N_1+0.25 N_2) \eta _B)\\+3 (p r (-12.57 B D N_2+C (24 \pi  N_1+18.87 N_2+12.57 A N_2)) v_r+C B_{\phi }^2 (3 r N_1 v_r+N_2 \eta _B))\\(r \alpha B_r B_{\phi }+r B_r^2 N_4+\rho  v_r (-12.57 r N_9 v_r+12.57 N_8 \eta _B+12.57 N_9 \eta _B)))\\+\frac{1}{-2+r}C B_{\phi} (3 r N_1 v_r-4(N_1+0.25 N_2) \eta _B) (H^2 r \alpha  B_r ((-2+3 r) B_{\phi }^2+8 \pi  (p (2+r)-2 (-2+r)\rho  v_r^2))\\+2 (-2+r) (-8 H^2 \pi  r \rho  B_r v_r v_{\phi }+B_{\phi } (H^2 r B_r^2 N_4+H r^2 B_r B_z N_6+4 \pi  \rho  (H^2 N_9+r^2 N_{13}) v_r \eta _B))))\\+\frac{1}{-2+r}H^2 (r (4 C \alpha  B_r B_{\phi } (25.13 p+B_{\phi}^2+25.13 \rho  v_r^2) (-0.75 r N_1 v_r+(N_1+0.25 N_2) \eta_B)\\+(((75.40\, +50.27 A) C-50.27 B D) p+3 C B_{\phi }^2) N_1 v_r (r \alpha  B_r B_{\phi }+r B_r^2 N_4+\rho  v_r (-12.57 r N_9 v_r+12.57 N_8 \eta _B+12.57 N_9 \eta _B)))\\-2 (12 C B_{\phi } (r \alpha  B_r (29.32 p+B_{\phi }^2+4.19 \rho  v_r^2)+B_{\phi } (-0.5 r B_r^2 N_5+\rho  N_8 v_r (6.28 r v_r-6.28 \eta_B)))\\(-0.75 r N_1 v_r+(N_1+0.25 N_2) \eta _B)+3 (p r (-12.57 B D N_2+C (24 \pi  N_1+18.87 N_2+12.57 A N_2)) v_r\\+C B_{\phi }^2 (3 r N_1 v_r+N_2 \eta _B)) (r \alpha  B_r B_{\phi }+r B_r^2 N_4+\rho  v_r (-12.57 r N_9 v_r+12.57 N_8 \eta _B+12.57 N_9 \eta_B))))\\(-8 (-1.5+r) B_{\phi }^2(r v_r-\eta _B)+2 (r v_r (p (201.06\, -125.66 r)-(-2+r) (B_z^2 N_3+4 \pi  \rho (v_{\phi }^2-r F'_r)))\\+\eta_B (p (-201.06+125.66 r)+(-2+r) (2 B_z^2 N_3+4 \pi  \rho  (v_{\phi }^2-r F'_r))))))),    \\
\end{multline}
and,
\begin{multline}\label{eq:denominator}
\mathcal{D}=(2 H^2 r ((150.80 p+6 B_{\phi }^2) (r v_r-\eta _B) (-37.70 C r \rho  B_{\phi } (2 \alpha  B_r+ B_{\phi } N_9) v_r^2 (r N_1 v_r-1.33 N_1 \eta _B-0.33 N_2 \eta _B)\\-3 (p r (-12.57 B D N_2+C (24 \pi  N_1+18.85 N_2+12.57 A N_2)) v_r+C B_{\phi }^2 (3 r N_1 v_r+1 N_2 \eta _B)) \\(r \alpha  B_r B_{\phi }+r B_r^2 N_4+\rho  v_r (-12.57 r N_9 v_r+12.57 N_8 \eta _B+12.57 N_9 \eta _B))\\+C B_{\phi } (3 r N_1 v_r-4 (N_1+0.25 N_2) \eta _B) (3 r \alpha  B_r (29.32 p+B_{\phi }^2+4.19 \rho  v_r^2)-1.5 B_{\phi } (r B_r^2 N_5+4 \pi  \rho  N_8 v_r (-r v_r+\eta _B))))\\+(r v_r (-87.97 p-B_z^2 N_3+4 \pi  \rho  v_r^2)+(87.97 p-4 \pi  \rho  v_r^2) \eta _B+3.50 B_{\phi }^2 (-r v_r+\eta _B))\\(r (4 C \alpha  B_r B_{\phi } (25.13 p+B_{\phi }^2+25.13 \rho  v_r^2) (-0.75 r N_1 v_r+(N_1+0.25 N_2) \eta _B)+\\ (((75.40\, +50.27 A) C-50.27 B D) p+3 C B_{\phi }^2) N_1 v_r (r \alpha  B_r B_{\phi }+r B_r^2 N_4+\rho  v_r (-12.57 r N_9 v_r+12.57 N_8 \eta _B+12.57 N_9 \eta _B)))\\-2 (12 C B_{\phi } (r \alpha  B_r (29.32 p+B_{\phi }^2+4.19 \rho  v_r^2)+B_{\phi } (-0.5 r B_r^2 N_5+\rho  N_8 v_r (6.28 r v_r-6.28 \eta _B)))\\ (-0.75 r N_1 v_r+(N_1+0.25 N_2) \eta _B)+3 (p r (-12.57 B D N_2+C (24 \pi  N_1+18.85 N_2+12.57 A N_2)) v_r\\+C B_{\phi }^2 (3 r N_1 v_r+N_2 \eta _B)) (r \alpha  B_r B_{\phi }+r B_r^2 N_4+\rho  v_r (-12.57 r N_9 v_r+12.57 N_8 \eta _B+12.57 N_9 \eta _B)))))).\\
\end{multline}
\bsp	
\label{lastpage}
\end{document}